\documentclass[twocolumn]{aastex63}

\pdfoutput=1 
\usepackage{amsmath,amstext,color}
\usepackage{gensymb}

\def\swift{{\it Swift}}

\def\grb{GRB\,161104A}

\shorttitle{GRB 161104a}
\shortauthors{Nugent et. al.}

\begin{document}

\title{The distant, galaxy cluster environment of the short GRB\,161104A at $z\sim 0.8$ and a comparison to the short GRB host population}

\newcommand{\NU}{\affiliation{Center for Interdisciplinary Exploration and Research in Astrophysics and Department of Physics and Astronomy, Northwestern University, 2145 Sheridan Road, Evanston, IL 60208-3112, USA}}
\newcommand{\CfA}{\affiliation{Center for Astrophysics-Harvard \& Smithsonian, 60 Garden Street, Cambridge, MA 02138-1516, USA}}
\newcommand{\OU}{\affiliation{Astrophysical Institute, Department of Physics and Astronomy, 251B Clippinger Lab, Ohio University, Athens, OH 45701, USA}}
\newcommand{\Carnegie}{\affiliation{Observatories of the Carnegie Institute for Science, 813 Santa Barbara Street, Pasadena, CA 91101-1232, USA}}
\newcommand{\Birmingham}{\affiliation{Birmingham Institute for Gravitational Wave Astronomy and School of Physics and Astronomy, University of Birmingham, Birmingham B15 2TT, UK}}
\newcommand{\UCB}{\affiliation{Department of Astronomy and Theoretical Astrophysics Center, University of California Berkeley, Berkeley, CA 94720}}
\newcommand{\FermiLab}{\affiliation{Cosmic Physics Center, Fermi National Accelerator Laboratory, PO Box 500, Batavia, IL 60510-5011, USA}}
\newcommand{\PSUAA}{\affiliation{Department of Astronomy and Astrophysics, The Pennsylvania State University, 525 Davey Laboratory, University Park, PA 16802, USA}}
\newcommand{\Purdue}{\affiliation{Purdue University, 
Department of Physics and Astronomy, 525 Northwestern Avenue, West Lafayette, IN 47907, USA}}
\author[0000-0002-2028-9329]{A.~E.~Nugent}
\NU

\author[0000-0002-7374-935X]{W.~Fong}
\NU

\author[0000-0002-9363-8606]{Y.~Dong}
\NU\Purdue

\author[0000-0002-6011-0530]{A. Palmese}
\FermiLab

\author[0000-0001-6755-1315]{J.~Leja}
\PSUAA

\author[0000-0003-3937-0618]{A.~Rouco~Escorial}
\NU

\author[0000-0003-0526-2248]{P.~K.~Blanchard}
\NU

\author[0000-0001-8340-3486]{K.~Paterson}
\NU

\author[0000-0002-7706-5668]{R.~Chornock}
\OU
\NU

\author[0000-0002-0048-2586]{A. Monson} 
\Carnegie

\author[0000-0002-2555-3192]{M. Nicholl} 
\Birmingham

\author[0000-0002-9392-9681]{E. Berger} 
\CfA

\begin{abstract}
We present optical observations of the {\it Swift} short-duration gamma-ray burst (GRB) \grb\ and its host galaxy at $z=0.793 \pm 0.003$. We model the multiband photometry and spectroscopy with the stellar population inference code \texttt{Prospector}, and explore the posterior using nested sampling. We find a mass-weighted age of $t_m = 2.12^{+0.23}_{-0.21}$~Gyr, stellar mass of $\log{(M/M_\odot)} = 10.21 \pm 0.04$, metallicity of $\log{(Z/Z_\odot)} = 0.08^{+0.05}_{-0.06}$, dust extinction of $A_V = 0.08^{+0.08}_{-0.05}$ mag, and a low star formation rate of $9.9 \times 10^{-2} M_\odot$~yr$^{-1}$. These properties, along with a prominent 4000~\AA\ break and optical absorption lines classify this host as an early-type, quiescent galaxy. Using Dark Energy Survey galaxy catalogues, we demonstrate that the host of GRB\,161104A resides on the outskirts of a galaxy cluster at $z\approx 0.8$, situated $\approx 1$~Mpc from the likely brightest cluster galaxy. We also present new modeling for 20 additional short GRB hosts ($\approx33\%$ of which are early-type galaxies), finding population medians of $\log(M/M_\odot) = 9.94^{+0.88}_{-0.98}$ and $t_m = 1.07^{+1.98}_{-0.67}$~Gyr ($68\%$ confidence). We further find that the host of \grb\ is more distant, less massive, and younger than the four other short GRB hosts known to be associated with galaxy clusters. Cluster short GRBs have faint afterglows, in the lower $\approx 11\%$ ($\approx 30\%$) of observed X-ray (optical) luminosities. We place a lower limit on the fraction of short GRBs in galaxy clusters versus those in the field of $\approx 5-13\%$, consistent with the fraction of stellar mass $\approx 10-20\%$ in galaxy clusters at redshifts $0.1 \leq z \leq 0.8$. Future studies that take advantage of wider-field and deeper cluster surveys are needed to understand the true rate of short GRBs in clusters and their effect on heavy-element enrichment in the intracluster medium.
\end{abstract}


\section{Introduction}
Short-duration $\gamma$-ray bursts (GRBs) have long been linked to a diverse set of stellar populations, ranging from galaxies with ongoing star formation to older, elliptical galaxies with deep upper limits on their star formation rates (SFRs; \citealt{Bloom2006,gcg+06,Fong2013, Berger_2014,dep19}). In the context of their progenitors, this diversity has been attributed to their origin from binary neutron star (BNS) and/or neutron star-black hole (NSBH) mergers, which are expected to have a broad range of delay times, in part governing the types of stellar populations in which short GRBs occur \citep{bpb+06,Nakar2006, Berger2007_HighZ, Zheng2007}. It is now known that the majority of short GRB hosts are indeed star-forming galaxies with moderate amounts of star formation of $\approx 0.1-1\,M_{\odot}$~yr$^{-1}$ \citep{Berger2009, Fong2013_Dem, Berger_2014}, with $\approx 1/3$ in early-type galaxies with limits on their star formation of $\lesssim 0.1\,M_{\odot}$~yr$^{-1}$.

A subset of short GRBs have likely associations with galaxy clusters, which represent the universe's largest gravitationally bound structures and compose $\sim 10-20\%$ of its stellar mass \citep{Fukugita1998, Eke2005}. Typically, galaxy clusters contain 10 or more galaxies and have total mass, including the dark matter halo, $> 10^{13.5}M_\odot$. The galaxy properties of those in clusters are distinct from those in the field: the frequency of early-type, older galaxies is higher than the field at similar redshifts ($\approx60\%$ in clusters and $\approx20\%$ in the field at $0.5 \leq z \leq 1.0$; \citealt{Tamburri2014, Hennig2017}), and the fraction of stellar mass within clusters is dominated by large, massive galaxies, which in turn affects their star formation histories (SFHs) and average stellar population ages \citep{LaraLopez2010, Peng2010, Lagan2018}. Overall, the amount of star formation in galaxy clusters is low compared to the field and contains a significantly older stellar population. Thus, identifying transients associated with galaxy clusters enhances the populations that require long delay times from formation to explosion, lending crucial insight to their progenitors and formation timescales.

The relationships between transients and their discovery in galaxy clusters have been useful in lending clues to their progenitors. For instance, the discovery of some calcium-strong transients in old galaxy cluster environments points to an old stellar progenitor for at least a fraction of these systems \citep{Lunnan2017, Frohmaier2018}. Type Ia supernovae (SNe Ia) have higher rates in both early-type field and cluster galaxies than Type Ib/c and II SNe; this is commensurate with their white dwarf (single- or double-degenerate) progenitors and massive star origins, respectively \citep{SNRateGC,Sand2008}.

For both long and short GRBs, a few studies have focused on events discovered with the Burst and Source Transient Experiment (BATSE) and The Imaging Compton Telescope (COMPTEL), which provided a large sample of degree-scale localizations. These studies cross-correlated GRB positions with available galaxy cluster catalogues. For instance, \citet{Marani1997} analyzed $\sim\!100$ BATSE and COMPTEL GRBs, finding very little correlation with clusters. \cite{Ghirlanda2006} reported weak correlations between short GRBs and galaxy clusters and additionally found that long GRBs have no correlation with clusters. \cite{Tanvir2005} cross-correlated BATSE short GRBs with nearby galaxies, including many in local clusters, finding positive correlations that become stronger when limiting to only early-type galaxies.

The launch of the Neil Gehrels {\it Swift} Observatory in 2004 \citep{Gehrels2004} has enabled well-localized GRB positions and subsequent host galaxy associations. Among the population of $\approx 130$ {\it Swift} short GRBs \citep{Lien2016}, $\lesssim 1/3$ have been robustly associated with host galaxies, and only three short GRBs have been reported as associated with galaxy clusters \citep{Prochaska2006, Berger2007}; where all three are with massive, quiescent, early-type galaxies. Given early associations of short GRBs with massive quiescent galaxies \citep{Berger_050724, Bloom2006, Gehrels2006, Prochaska2006, GRB050813, Bloom2007}, and the scaling of globular cluster frequency with stellar mass, it was originally thought that $10-30\%$ of short GRBs could be dynamically formed in globular clusters \citep{Grindlay2006}. However, more recent theoretical and observational studies have shown that globular clusters cannot contribute significantly to the fraction of BNS and NSBH mergers \citep{Belczynski2018, Fong2019, Lamb2019, Ye2020}. It is nonetheless expected that the rate of short GRBs in clusters matches the fraction of stellar mass in galaxy clusters. Finally, given that BNS mergers are in part responsible for $r$-process enrichment (e.g., \citealt{elp89,ros05,gbj11,kra+12,cbk+17, Drout2017, Kasen2017, Pian2017,met19}), the rate of short GRBs in clusters can be used to trace heavy-element enrichment in the ICM, akin to studies focused on reconciling metal enrichment of the ICM through studies of cluster Ca-strong transients and SNe Ia \citep{mkk14}.

Here we present observations of the {\it Swift} short GRB, \grb, and the identification of its large-scale environment as a galaxy cluster at a median redshift of $z\approx 0.79$. This event adds to a small subset of short GRBs known to be associated with galaxy clusters and is the highest-redshift cluster association to date. In Section~\ref{sec:obs}, we present the observational data of the \grb\ and the galaxies in the immediate vicinity. We discuss the large-scale environment of the region containing \grb\ and its cluster association using galaxy catalogues in Section \ref{sec:cluster}. We describe our stellar population fitting of the host of \grb\ and several surrounding galaxies and present a uniform reanalysis of a sample of 20 short GRB hosts in Section \ref{sec:stellarpopmodel}. We also identify the large-scale environment of another short GRB, GRB\,090515, as a galaxy cluster at $z\approx 0.4$. Finally, we compare \grb\ to the other known short GRBs in galaxy clusters, the short GRB host population, and other transients discovered in galaxy clusters in Section \ref{sec:discussion}. Unless otherwise noted, magnitudes are in the AB system and uncertainties correspond to $1\sigma$ confidence. We employ a flat $\Lambda$CDM cosmology of Hubble constant $H_{0}$ = 69.6 ${\rm km \, s}^{-1} \, {\rm Mpc}^{-1}$, matter density $\Omega_{M}$ = 0.286, and cosmological constant $\Omega_{\rm vac}$ = 0.714 \citep{Bennett2014}.

\begin{deluxetable*}{lccccccccc}[!t]
\tabletypesize{\footnotesize}
\tablecolumns{10}
\tablewidth{0pc}
\tablecaption{GRB 161104A Host Galaxy Photometry 
\label{tab:phot}}
\tablehead{
\colhead {Date}	 &
\colhead{Facility} &
\colhead {Instrument}  &
\colhead{Exposures} &
\colhead {Band}	 &
\colhead {G1 (Host)}	&
\colhead {G2}		    &
\colhead {G3}		    &
\colhead {G4}		    &
\colhead {G5}		    \\
\colhead {(UT)}	 &
\colhead{} &
\colhead {}  &
\colhead{(s)} &
\colhead {}	 &
\colhead {(AB mag)}		    &
\colhead {(AB mag)}		    &
\colhead {(AB mag)}		    &
\colhead {(AB mag)}		    &
\colhead {(AB mag)}		      
}
\startdata
2018 Jan 7 & Magellan/Baade & IMACS & $5 \times 420$ & $g$ & $25.44 \pm 0.25$ & $>25.5$ & $>25.5$ & $24.63 \pm 0.08$ & $24.31 \pm 0.12$ \\
2016 Nov 6 & Gemini-South & GMOS & $6 \times 120$ & $r$ & $23.86 \pm 0.11$ & $25.20 \pm 0.19$ & $24.51 \pm 0.07$ & $24.16 \pm 0.06$ & $22.56 \pm 0.03$ \\
2016 Nov 7 & Magellan/Baade & IMACS & $6 \times 360$ & $r$ & $23.81 \pm 0.10$ & $24.95 \pm 0.16$ & $24.72 \pm 0.09$ & $24.12 \pm 0.07$ & $22.79 \pm 0.05$ \\ 
2016 Nov 7 & Magellan/Baade & IMACS & $6\times240$ & $i$ & $22.72 \pm 0.06$ & $24.96 \pm 0.20$ & $23.48 \pm 0.07$ & $23.86 \pm 0.1$ & $21.29 \pm 0.05$ \\
2018 Jan 7 & Magellan/Baade & IMACS & $11\times180$  & $z$ & $22.14 \pm 0.07$ & $23.98 \pm 0.17$ & $23.00 \pm 0.07$ & $23.52 \pm 0.20$ & $21.00 \pm 0.06$ \\
2016 Nov 8 & Magellan/Baade & Fourstar & $33\times61.13$ & $J$ & $21.56 \pm 0.04$ & $24.41 \pm 0.49$ & $21.97 \pm 0.08$ & $22.91 \pm 0.13$ & $19.74 \pm 0.04$
\enddata
\tablecomments{The most probable host galaxy of \grb\ is Galaxy ``G1''. All magnitude values have been corrected for Galactic extinction in the direction of the GRB \citep{sf11}. The values that we use for each filter are $A_g = 0.054$~mag, $A_r = 0.037$~mag, $A_i = 0.028$~mag, $A_z = 0.021$~mag, and $A_J = 0.012$~mag.}
\end{deluxetable*} 

\section{Observations}
\label{sec:obs}
\subsection{Discovery of GRB\,161104A}
\label{sec:grbinfo}
The Neil Gehrels $\swift$ Observatory's \citep{Gehrels2004} Burst Alert Telescope (BAT) triggered on \grb\ at 09:42:26 UT and measured a single-peaked light curve with duration $T_{90} = 0.10 \pm 0.02$~s (15-350~keV), thereby qualifying it as a short GRB \citep{GCN20123,GCN20129}. The fluence was measured to be $f_{\gamma} = (3.1 \pm 0.5) \times 10^{-8}$ erg~cm$^{-2}$ in the 15-150 keV band.  At the time of the trigger, BAT localized the burst to R.A. = $05^{\text h} 11^{\text m} 31^{\text s}$, decl. = $-51 \degree 27' 07''$ (J2000), with a $3'$ radius and 90$\%$ containment. Based on the data taken with the Photon Counting mode on the X-ray Telescope (XRT), this localization was later improved to a $3''.2$ radius at R.A. = $05^{\text h} 11^{\text m} 34.5^{\text s}$, decl. =  $-51\degree 27' 36.4''$, with 90$\%$ confidence \citep{Evans, GCN20123}. A single X-ray data point was detected at $\delta t \approx 10^2$~s (where $\delta t$ is the time since the BAT trigger) but faded below the XRT detection limit by $\delta t \approx 10^4$~s. Approximately $67$~s after the BAT trigger, the Swift Ultraviolet/Optical Telescope (UVOT) observed the position of \grb\ with the {\it white} filter, finding no optical afterglow to $> 20.8$ mag \citep{GCN20123}.

Further ground-based follow-up observations were taken with the Gamma-Ray Burst Optical/Near-Infrared Detector (GROND) on the MPG telescope at the European Southern Observatory \citep{GCN20132} and Gemini-South \citep{GCN20137}. GROND took simultaneous $griz$-$JHK$ images at $\delta t = 18$~hr and Gemini-South took $r$ band imaging at $\delta t \sim 17$~hr. Both observations identified four optical sources within or around the XRT position, which will be discussed further in Section~\ref{sec:photometry}. 

Additionally, there was X-ray follow-up at $\delta t = 364$~ks with the Chandra X-ray Observatory \citep{Weisskopf2000} that reported an upper limit of $F_x < 4.5\times10^{-15}$~erg~s$^{-1}$~cm$^{-2}$ \citep{GCN20166}.

\subsection{Optical and Near-infrared Observations}
\label{sec:photometry}
\begin{figure*}
\includegraphics[width=1.0\textwidth]{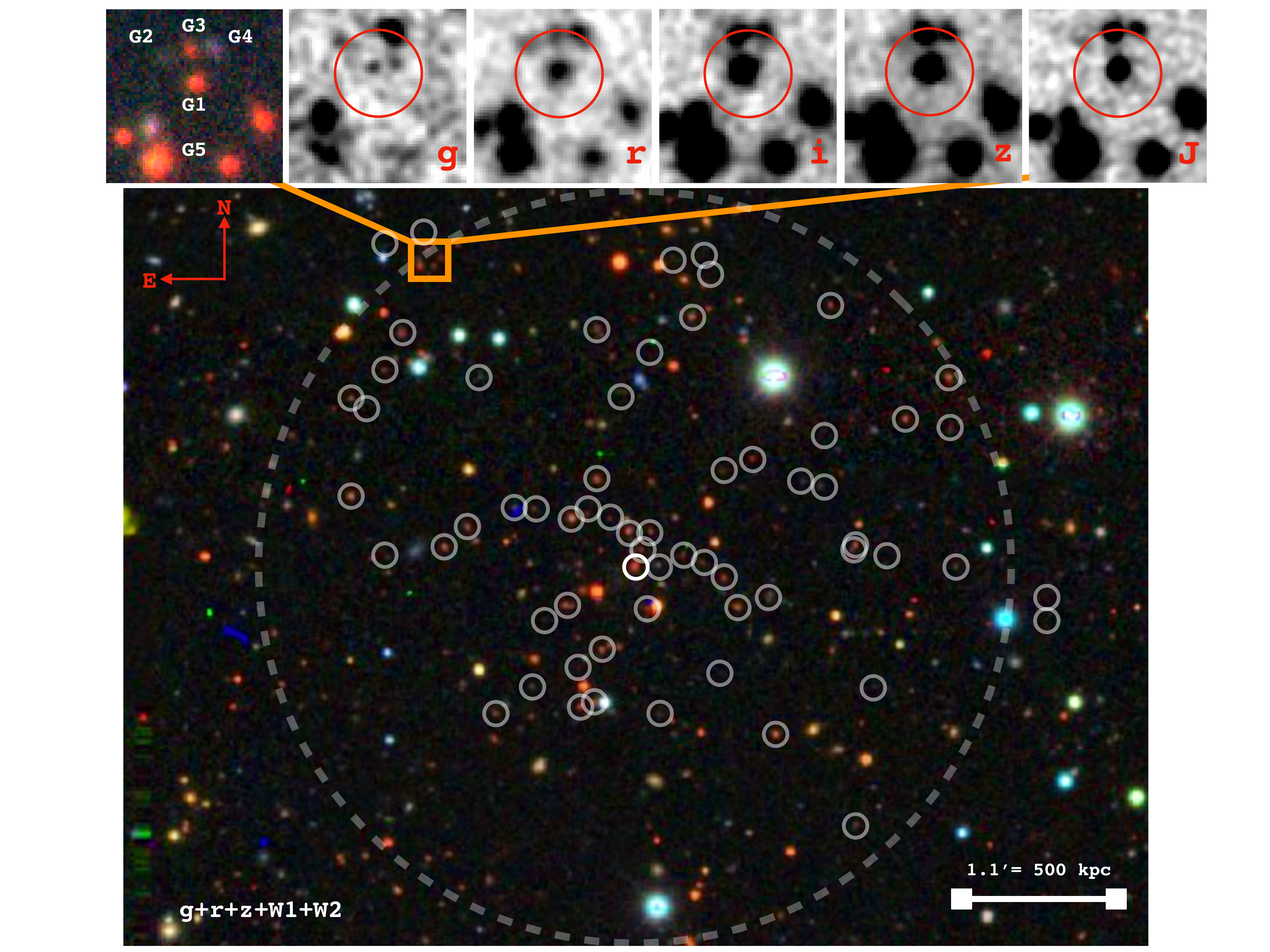}
\vspace{-0.1in}
\caption{Wide-field composite image of \grb, the associated cluster, and surrounding galaxies \citep{LegacySurveyDr8}. The cluster radius is outlined by the grey dashed, circle, cluster members are circled in grey, and the BCG is circled in white. Cluster members with $r$ band magnitude $> 23$ are not visible in the composite, though their positions are circled. At the top, going from left to right: a zoom-in of a Magellan/IMACS color composite ($grz$) image of G1, made with \texttt{AstroImageJ}
\citep{AstroImageJ}, encompassed by the XRT position (red circle, 90$\%$ containment), and G2, G3, and G4 on the outskirts, $g$ band, $r$ band, $i$ band, $z$ band, $J$ band from Magellan. G5, which is outside the XRT position, is also labeled. The red colors and source density of G1, G2, G3, and G5 indicate a cluster or group environment, while the blue color of G4 suggests that this galaxy originates at a different redshift.}
\label{composite}
\end{figure*}

We retrieved available imaging of the location of \grb\ taken with the Gemini Multi-Object Spectrograph (GMOS) mounted on the $8$~m Gemini-South telescope (PI Troja; GS-2016B-Q-28) and first reported in \cite{GCN20137}. The GMOS observations consist of $6\times120$~s of $r$ band imaging taken at a midtime of $\delta t = 17.3$~hr. We create bias and flat-field frames using associated calibrations in the Gemini archive, and apply these calibrations to the science images using the {\tt gemini} package in IRAF. We produce a median-combined image and perform astrometry with stars in common with the Two Micron All Sky Survey (2MASS) catalogue. We identify three extended sources coincident with the XRT position, only one of which is fully encompassed by the XRT position. This source is clearly a galaxy, first reported in \citealt{GCN20132} (Figure \ref{composite}).

We acquired further, deeper imaging on 2016 November 7 UT at $\delta t = 2.77$~days (PI: Fong) with the Inamori-Magellan Areal Camera and Spectrograph (IMACS) instrument on the $6.5$m Magellan-Baade telescope in better conditions than the initial Gemini imaging. We took $6\times120$~s in the $r$ band and $6\times240$~s in the $i$ band. We reduced and stacked the data in the same manner as the Gemini data, using standard packages in the {\tt ccdred} package in IRAF. We identify the same three sources detected by Gemini and one additional source along the edge of the XRT position. Hereafter we refer to these sources as G1 (source fully within the XRT position), G2, G3, and G4 (Figure \ref{composite}).

To assess the presence of an afterglow related to any of the sources, we perform image subtraction between the Gemini and IMACS $r$ band epochs using the {\tt HOTPANTS} software package \citep{bec15}. The lack of any sources in or around the XRT position in the residual image enables us to place a limit on the afterglow emission. We use the IRAF/{\tt phot} package \citep{iraf1, iraf2} to perform aperture photometry on faint sources in the Gemini $r$ band observations and place a $3\sigma$ limit of $r\gtrsim 25.4$~mag at $\delta t = 17.3$~hr, calibrated to a standard star field at similar air mass.

To better characterize G1-G4, we obtained $J$ band imaging with Fourstar on the Magellan-Baade telescope on 2016 November 8 UT ($\delta t \approx 3.75$ days) reduced with a custom pipeline, as well as late-time $g$ and $z$ band imaging with IMACS on 2018 January 7 UT. For all filters, we perform astrometry using the Astrometry.net software, which uses sources in common with the USNO-B and 2MASS catalogues for absolute astrometry \citep{Lang_2010}.

We perform photometry of all four sources, as well as a nearby galaxy, G5, which is serendipitously covered by our spectroscopy (Section~\ref{sec:spectroscopy}). To determine the zero-point of the images, we use a standard-star field at a similar air mass for the optical imaging and the 2MASS catalogue for $J$ band, and then convert to the AB system. We then use IRAF/{\tt phot}, defaulting to an aperture of $2.5\times\theta_{\text{FWHM}}$. Since the field is crowded, we select smaller apertures in some cases to avoid contamination from nearby objects. We correct the magnitudes of the sources for Galactic extinction, $A_{\lambda}$ \citep{sf11}. In Table \ref{tab:phot}, we present the details of the imaging observations and photometry of G1-G5. Figure \ref{composite} shows a color composite of the larger field of view of \grb, as well as the positions of G1-G5. We calculate G1 to be at a position of R.A. = $5^{\text h} 11^{\text m} 34.47^{\text s}$, decl. = $-51\degree 27' 36.29"$ (J2000).

\subsection{Spectroscopy}
\label{sec:spectroscopy}
On 8 Nov 2016 UT, we obtained $3\times1800$~s of spectroscopy with Magellan/IMACS using a $0''.7$ slit and the 200-line grism in the {\it f}/2 camera (first reported by \citealt{GCN20168}). The slit passed through G1, G4, and serendipitously through another galaxy, G5. The spectrum covered the optical wavelength range of $\sim4500-10050$ \AA.  Basic two-dimensional image processing tasks and spectral extraction were performed in IRAF, while a flux calibration was applied using custom IDL routines. The spectrum of G1 exhibits a red continuum, lacks emission lines, has a discernible 4000 \AA\ break at $\sim 7172$ \AA, and has distinguishable Ca II H and K absorption lines. These features classify G1 as an early-type, quiescent galaxy at $z \sim 0.79$. To determine the precise redshift and uncertainty, we cross-correlated with a galaxy template at age 2.5 Gyr using the model described in \cite{BC2003} and find that $z = 0.793 \pm 0.003$. The spectrum of G4 exhibits a bluer continuum with a single emission line around 9780 \AA. Similar to G1, the spectrum of G5 exhibits clear early-type, quiescent galaxy features. Based on the identifications of Ca II H and K absorption lines and a clear 4000 \AA\ break, G5 also has a redshift of $z \sim 0.79$.

On 2017 February 2 UT, we took $3\times1800$~s dithered exposures of spectroscopy with the Low Dispersion Survey Spectrograph (LDSS) on the Magellan-Clay telescope, with the slit passing through G2, G3, and G4 (PI Berger). These spectra were also in the optical wavelength range, covering $\sim3800-10600$ \AA. We applied a bias correction and flat-field correction, aligned the dithered frames, and combined using standard tasks in the {\tt ccdred}, {\tt longslit}, and {\tt immatch} packages in IRAF \citep{iraf1, iraf2}. We identified and extracted the spectral traces with the {\tt apextract} package. We then applied a wavelength solution to the spectra using Ne-He-Ar arc lamps taken on the same night and a flux calibration using the standard star LTT 4364, with tasks from the {\tt onedspec} package. We also extracted error spectra using the co-added spectra with no background subtraction and divided by $\sqrt{N}$, where $N$ is the number of exposures. G3 exhibits a red continuum with no obvious features, but it is too faint for a meaningful extraction. We identified a tentative 4000 \AA\ break in the spectrum of G3 around 7000-7200 \AA, translating to a probable redshift around $0.75 < z < 0.80$; there are no other apparent emission or absorption lines. We identified the same emission line as previously discussed in the spectrum of G4; however, no other lines are evident.

\section{Large-scale Environment}
\label{sec:cluster}

\subsection{Putative Host Galaxy}
\label{sec:putative}
\begin{figure}
\centering
\includegraphics[width=0.475\textwidth]{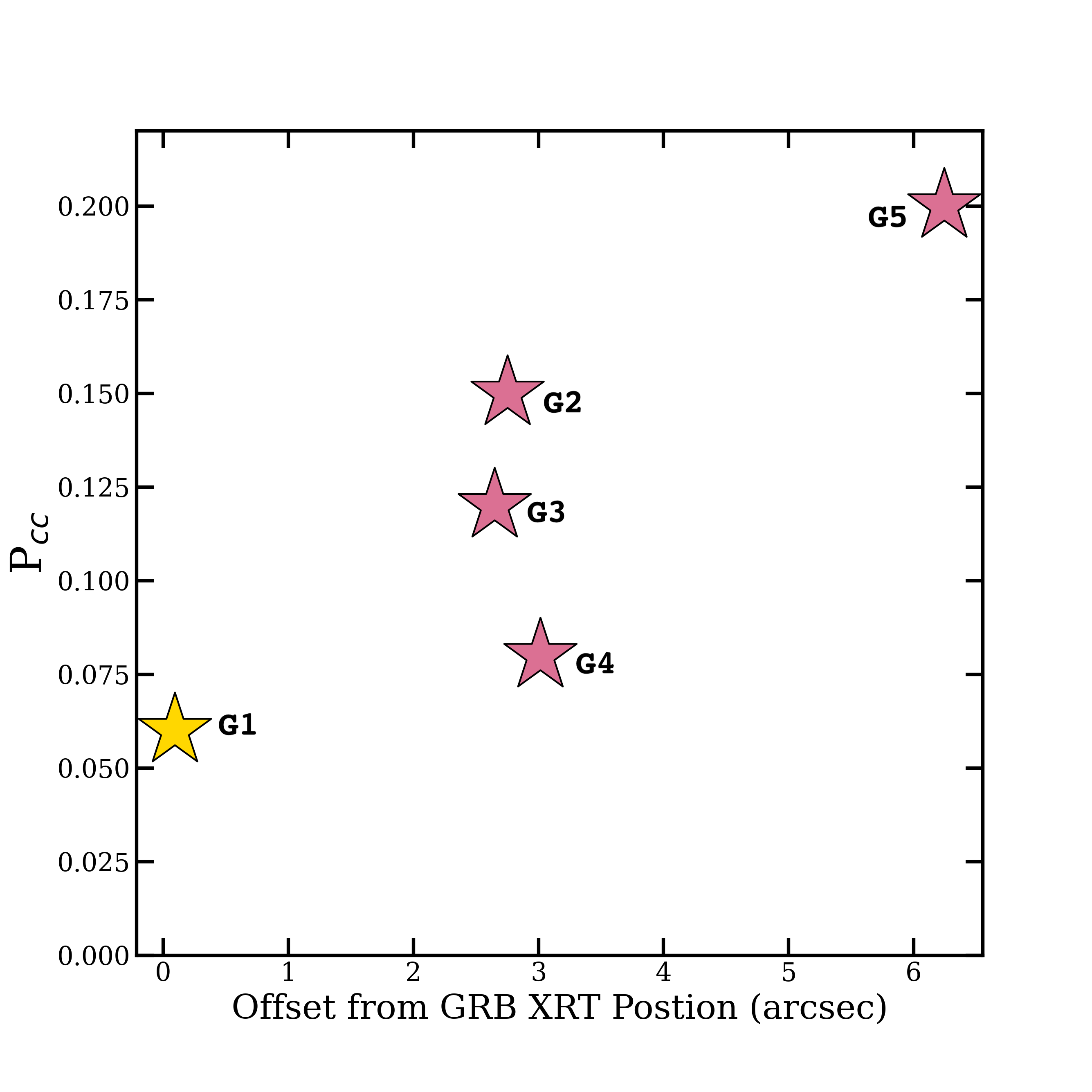}
\vspace{-0.3in}
\caption{Probability of chance coincidence ($P_{cc}$) for each galaxy near the XRT position of \grb\ versus offset from the \grb. G1 is yellow, whereas G2-G5 are pink. G1 is the only galaxy fully encompassed by the XRT position and is the brightest of the galaxies coincident with XRT position. This results in the lowest $P_{cc}$ for G1, making it the putative host.}
\label{fig:Pcc}
\end{figure}

To quantify the likelihood that each of the sources G1-G5 is the putative host of \grb, we calculate the probability of chance coincidence, $P_{cc}$, at a given distance $R_i$, and apparent optical magnitude ($m$). In this method, a lower value of $P_{cc}$ translates to a larger probability of being the host galaxy \citep{bkd02}. The value of $R_i$ is taken to be the maximum of $\left[2r_e, \sqrt{\delta R + 4r_e}, 3\sqrt{\sigma_{\rm tie}^2+\sigma_{\rm GRB}^2}\right]$ \citep{bkd02,bbf16}, where $r_e$ is the half-light radius of the galaxy and $\delta R$ is the angular separation between the afterglow position and galaxy. For G1-G4, the final term clearly dominates owing to the relatively large XRT positional uncertainty of $\sigma_{\rm GRB} = 2''.25$ (converted to $1\sigma$ confidence). Thus, taking into account the $1\sigma$ astrometric tie uncertainty of $\sigma_{\rm tie}=0''.19$, we use $R_i=2''.26$ and our measured $r$-band magnitudes (Table~\ref{tab:phot}) to calculate the values of $P_{cc}$. We plot the $P_{cc}$ value versus galaxy offset from the GRB XRT position in Figure \ref{fig:Pcc}. We find that G1 has the lowest value of $P_{cc} = 0.06$, G4 is the next most probable galaxy with $0.08$, and G2 and G3 have values of $0.15$ and $0.12$, respectively.

\begin{figure}
\includegraphics[width=0.475\textwidth]{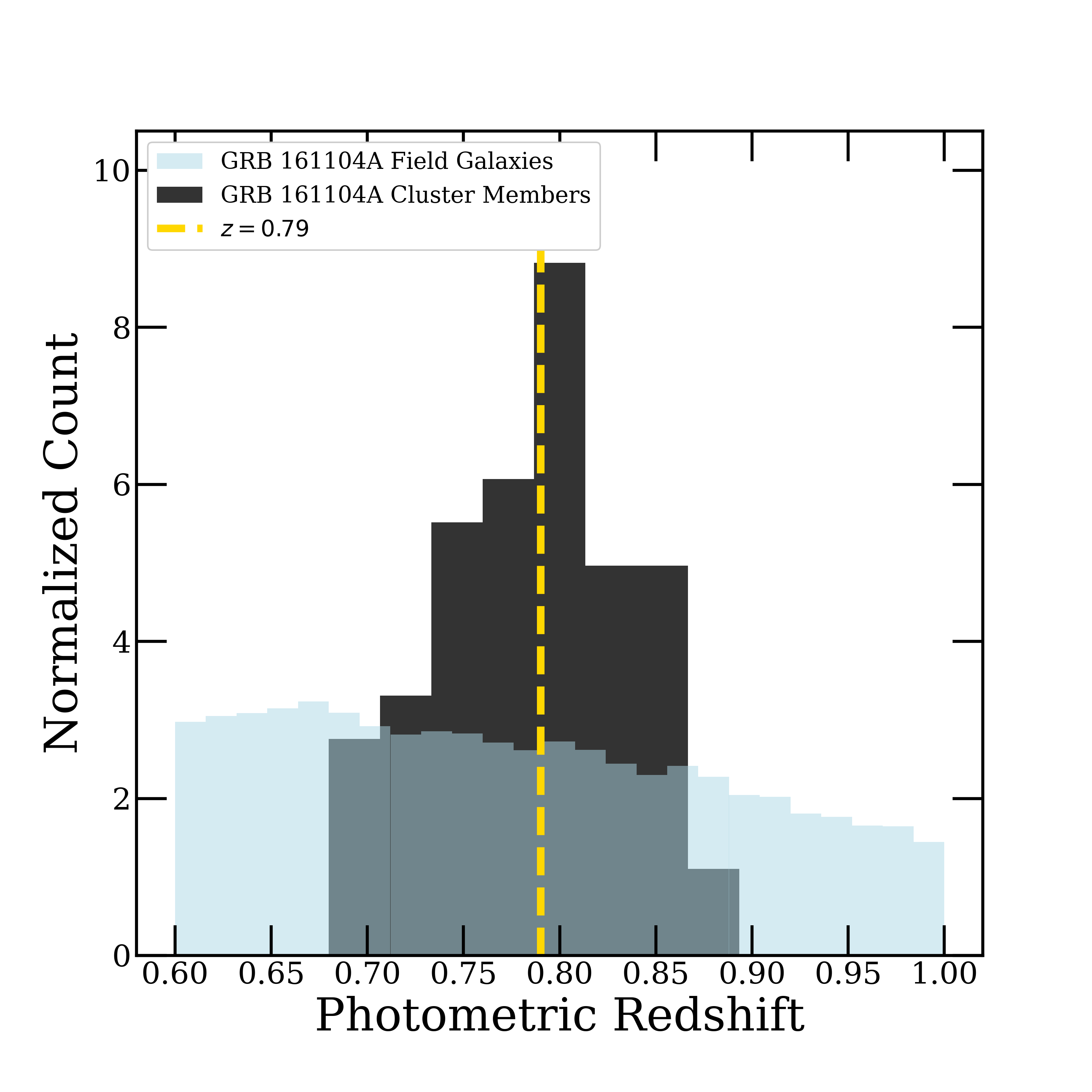}
\caption{Comparison of the photometric redshift distributions of the putative cluster of \grb\ (black) to all of the galaxies in the DES Year 1 GOLD dataset surrounding \grb\ in a $\sim 1.5$ deg$^2$ area (blue). Both distributions have been normalized to have an area of unity. For reference, we mark the spectroscopic redshift of the putative host galaxy of \grb, G1, at $z=0.793$ (yellow dashed line). We find that the redshift distribution of the putative cluster members in the field of \grb\ peaks at $z \sim 0.79$, which is consistent with the redshift of G1, solidifying the origin of \grb\ from a galaxy cluster.}
\label{fig:red_hist}
\end{figure}

\begin{figure*}
\includegraphics[width=1.0\textwidth]{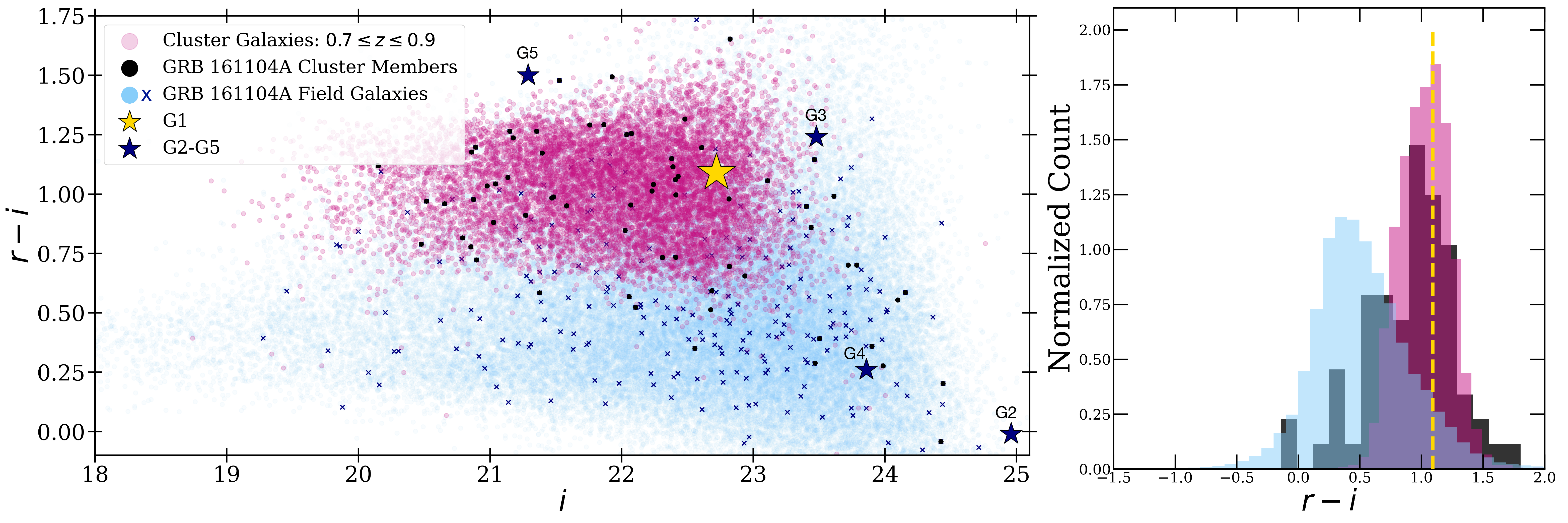}
\caption{{\it Left:} color-magnitude diagram of the putative cluster of \grb\ (black) compared to all the DES Year 1 galaxies surrounding \grb\ within a 1.5 deg$^2$ or 2150 Mpc radius (blue) and to DES SV redMaPPer cluster galaxies at $0.7 \leq z \leq 0.9$ (pink). We also highlight the galaxies within $130''$ of the BCG for which we do not include as ``cluster members'' (dark-blue crosses), showing that these more proximal field galaxies trace a similar part of the color-magnitude space to the field galaxies selected from a larger radius. The stars indicate the galaxies studied in this paper, where the host, G1, is yellow and G2-G5 are dark-blue. {\it Right:} histogram of the $r$-$i$ colors, following the same color scheme as in the left panel. We normalize the histograms such that the density of each is 1.  The dashed yellow line represents the $r$-$i$ color of G1. The galaxies in the putative cluster, as well as G1, have remarkably similar colors to those in the redMaPPer sample and are noticeably redder than the other galaxies in the field. We also find additional confirmation that G1, G3, and G5 are likely cluster members, given the similarity of colors between them, the known clusters, and the putative cluster of \grb.}
\label{fig:CMD}
\end{figure*}

We also calculate the probabilities for all extended sources within $15'$ of the XRT position, finding that G5 has the next-lowest value of $P_{cc}< 0.21$, while the remaining sources have values that exceed 0.5. Taken together, we find that G1 is the most probable host galaxy of \grb. However, most short GRB host associations have been made based on values of $P_{cc} \lesssim 0.05$, given the large observed offsets from of short GRBs from their host galaxies \citep{Fong2013_Dem}. Given the similar $P_{cc}$ value derived for G4, we cannot immediately discount an origin from G4 (Figure \ref{composite}).

\subsection{Cluster Membership}
\label{sec:clustermem}
As seen in Figure \ref{composite}, there are many galaxies in the vicinity of \grb, several of which exhibit similar colors. To assess their membership in a cluster or group, as well as the large-scale environment of \grb, we analyze a $\sim 1.5$ deg$^2$ area around G1 using the Dark Energy Survey (DES; \citealt{descollaboration,flaugher,2016MNRAS.460.1270D}) Year 1 GOLD dataset of galaxies \citep{DESY1}, which includes high-precision photometric data, including extinction-corrected multiband photometry, star-galaxy classification \citep{2018MNRAS.481.5451S}, and accurate photometric redshifts ($z_{\text {phot}}$; \citealt{Hoyle_2018}). We note that although there are X-ray data with Chandra and XMM, they are not sufficiently sensitive to place meaningful constraints on diffuse X-ray emission that are typical of lower-redshift and massive galaxy clusters \citep{FJ1982, Sarazin1986}. Moreover, we cannot define the cluster using velocity dispersion, as we do not have spectroscopic data of all the galaxies in the region and such studies are extremely challenging. Thus, we rely on the well-established photometric red sequence technique to determine determine cluster membership \citep{Redmapper}.

We begin by filtering all of the galaxies in the region to only include those with $0.7 \leq z_{\text {phot}} \leq 0.9$, as the photometric redshifts have an uncertainty of at least $\pm 0.1$. We then select a $\sim 130''$ radius area around G1, as this corresponds to a typical cluster radius of $1$~Mpc at $z\approx0.8$. We determine that the brightest red galaxy in this region, i.e. the likely brightest cluster galaxy (BCG) lies at R.A. = $05^{\text{h}} 11^{\text{m}} 26.67^{\text{s}}$, decl. = $-51^{\degree} 29' 27.43''$, about 1 Mpc from G1, with a magnitude of $r=21.27$~mag. We then shift the center of the cluster to the BCG and resolve that the rest of the cluster members are the reddest galaxies within a 1 Mpc radius of the BCG. We find that 68 galaxies in this region fit these classifications, and we designate these as ``\grb\ cluster members.'' These galaxies are also marked in Figure \ref{composite}. We note that although we are able to locate G1 and G5 and include this in our filtering, the other galaxies surrounding \grb\ are too faint for the survey limits. Figure \ref{fig:red_hist} shows the distribution of redshifts in the cluster sample compared to all galaxies in the $\sim 1.5$ deg$^2$ region around \grb\ considered here. We see a clear peak at the redshift of the putative host galaxy ($z \approx 0.79$), while galaxies in the field (``\grb\ Field Galaxies'') span a wide range of redshifts as expected.

We next compare the colors and magnitudes of the galaxies in the putative cluster of \grb\ to those of known cluster members at similar redshifts. We use the DES Scientific Verification (SV) Red-Sequence Matched-Filter Probabilistic Percolation Cluster Finder (redMaPPer; \citealt{SDSS_Cluster}) catalogue of galaxy clusters \citep{Redmapper} to collect all galaxies in clusters with $0.7 \leq z_{\text {phot}} \leq 0.9$. Our filtered sample includes 158 clusters containing a total of 11,019 galaxies.  These galaxies are further compared with the $\sim 61,000$ galaxies in a region significantly wider than the cluster size (1.5 deg$^2$ in area, corresponding to $\approx 2510$~Mpc) surrounding \grb, which serve as a generic background/foreground galaxy population around the location of interest.

We determine the $r$-$i$ color for all galaxies in the three populations: \grb\ cluster members, field galaxies, and redMaPPer cluster galaxies. We select $r$-$i$ since the 4000 \AA\ break at $z\sim0.8$ would fall between these bands, a feature that is on average more pronounced for older cluster galaxies, and which helps to separate galaxy populations into red sequence and blue cloud. We show a color-magnitude diagram and a histogram of the colors for all three populations in Figure~\ref{fig:CMD}. Overall, we find that the majority of the galaxies in the \grb\ cluster member sample fall within the expected color-magnitude range for cluster galaxies at similar redshifts and are redder than the \grb\ field galaxy sample. Finally, we explore the field galaxies within $130''$ of the BCG, corresponding to $\sim 280$ galaxies, to ensure that the galaxies we have selected as noncluster members trace the same part of parameter space as the field galaxies selected from a larger region. Indeed, we find that they occupy the same region in the color-magnitude diagram as the wider-field galaxy selection (Figure~\ref{fig:CMD}), thus lending weight to our ``cluster members'' criteria. 

Focusing on the five galaxies in the vicinity of \grb, we find that G1, G3, and G5 exhibit similar redder colors both to galaxies in the putative cluster and to the redMaPPer known cluster sample, which supports our conclusion that they are all cluster members. We also note that G4 clearly exhibits bluer colors than the red sequence (see Section~\ref{sec:prospector}) and G2 is too faint to infer proper membership. Thus, we find that if \grb\ is associated with G1, then it likely originated from a galaxy cluster. 

To further assess \grb's association with the galaxy cluster, we find the probability that a random sight line intersects a galaxy cluster, by determining the full angular sky coverage occupied by clusters using the redMaPPer catalog (which is fairly complete to $z\sim 0.9$). Specifically, for each cluster in the redMaPPer sample, we calculate the angular extent on the sky assuming a 1~Mpc radius at the relevant redshift of each cluster, which gives the total square degrees occupied by redMaPPer clusters. We then use the total sky coverage of the survey ($\sim$116 deg$^2$) to calculate the chance of a sight line intersecting any cluster, assuming that the redMaPPer area surveyed is representative of the full sky. We find that for the full sample of clusters, the chance of intersecting any cluster is $\approx4.8\%$ ($\sim2.0\sigma$). However, given that the putative host cluster redshift of \grb\ is $z\sim0.8$, we recalculate this using only redMaPPer clusters at this redshift and find that the probability of intersecting a cluster at this redshift is significantly lower, at 0.6\% ($\sim2.9\sigma$). Thus, \grb\ has a low probability of being randomly aligned with a $z \sim 0.8$ cluster within our detection limits. We also modify the $P_{cc}$ method described in Section \ref{sec:putative} to account for the surface density of galaxies (in number arcsec$^{-2}$) in clusters from \cite{Redmapper} brighter than or equal to the $i$ band magnitude of G1 ($i\leq 22.72$). This modified equation would thus measure the probability of chance coincidence that the GRB is associated with a galaxy within a cluster, finding $P_{\rm cc} = 5.0 \times 10^{-4}$, which suggests that it is unlikely that \grb\ would occur in a cluster galaxy by chance.

\section{Stellar Population Modeling}
\label{sec:stellarpopmodel}

\subsection{Prospector}
\label{sec:prospector}
To determine the stellar population properties of G1 and the surrounding galaxies, we model their available data with the Python package \texttt{Prospector}~\citep{Leja_2017}.  \texttt{Prospector} is a stellar population inference code that applies a nested sampling fitting routine to the available observed photometry and spectroscopy of a galaxy to determine properties such as redshift ($z$), stellar mass ($M$), stellar population age, SFH, dust extinction ($A_V$), and metallicity ($Z$). These properties can either be set to a fixed value or varied over a specified range to determine the best-fit values. For each free parameter, \texttt{Prospector} returns a full posterior distribution, allowing for the determination of accurate uncertainties in individual parameters. In addition, \texttt{Prospector} utilizes \texttt{dynesty} \citep{Dynesty} to perform nested sampling, \texttt{Python-FSPS} (Flexible Stellar Population synthesis; \citealt{FSPS_2009, FSPS_2010}) to build its stellar population models, and also uses WMAP9 cosmology internally \citep{Hinshaw2013}.

For all of our stellar population models, we use a Chabrier initial mass function (IMF) \citep{Chabrier2003}, the Milky Way Extinction Law \citep{MilkyWay}, and a parametric delayed-$\tau$ SFH, where
\begin{equation*}
\text{SFR}(t) = M_F\times \Big[\int_0^t{te^{-t/\tau} dt}\Big]^{-1} \times te^{-t/\tau}, 
\end{equation*} 
with $M_F$ the total mass formed and the sampled mass metric, and $t = t_{\text{SF}}$, a free parameter that describes the look-back time at which star formation commences. Similar to the other stellar population properties, $\tau$ can also be set to a free parameter. The parameters $\tau$ and $t_{\text{SF}}$ can be used to find $t_m$, the mass-weighted stellar population age, through the equation 
\begin{equation*}
    t_m = t_{\text{SF}} - \frac{\int_0^{t_{\text{SF}}} t \times \text{SFR}(t) dt}{ \int_0^ {t_{\text{SF}}} \text{SFR}(t) dt}.
\end{equation*} 
The mass-weighted age is a more physically meaningful metric of age than the parameter $t_{\text{SF}}$, which simply measures the time at which star formation commenced \citep{Conroy2013SED}. We can further use $t_m$ and $M_F$ to find stellar mass in units with the approximation
\begin{equation*}
    M \approx M_F \times 10^{1.06 - 0.24\log{(t_m)} + 0.01*\log{(t_m)}^2}
\end{equation*} \citep{Leja2013}. Stellar mass is the preferred mass metric as opposed to total mass formed, as it measures the mass retained by the stellar population as it was observed. From here on, we only quote the mass-weighted ages ($t_m$) and stellar masses ($M$). Furthermore, we allow for nebular emission in all our fits to gauge the strength and location of spectral lines. For fits that include an observed spectrum, we add two additional parameters to our fits: a fixed parameter $n$, where $n$ is the order of the Chebyshev polynomial to fit the continuum of the observed spectrum, and a free parameter, $N_0$, which describes the normalization of the observed spectrum to the model spectrum continuum and should ideally converge to 1. If we have prior knowledge of the redshift of the galaxy and apply it in our fits, we set the maximum possible age to the age of the universe at that redshift. For galaxies with no known redshift, we set the maximum ages to the age of the universe at the minimum possible redshift. We determine the maximum age constraints using \cite{CosmologyCalc}, which also uses WMAP9 cosmology.

For this work, we find that the delayed-$\tau$ SFH is advantageous, as opposed to a nonparametric SFH, as it is directly comparable to most previous work in the field and enables uniform analysis for galaxies with only photometry \citep{Conroy2013SED, Leja2019}. We note that the major difference between full nonparametric and delayed-$\tau$ models for the SFH is in the age and mass estimates: as shown in \cite{Leja2019}, ages are systematically three to five times older, while masses are typically 25\%-100\% larger.

\begin{figure*}[t]
\makebox[\textwidth][c]{\includegraphics[width=\textwidth]{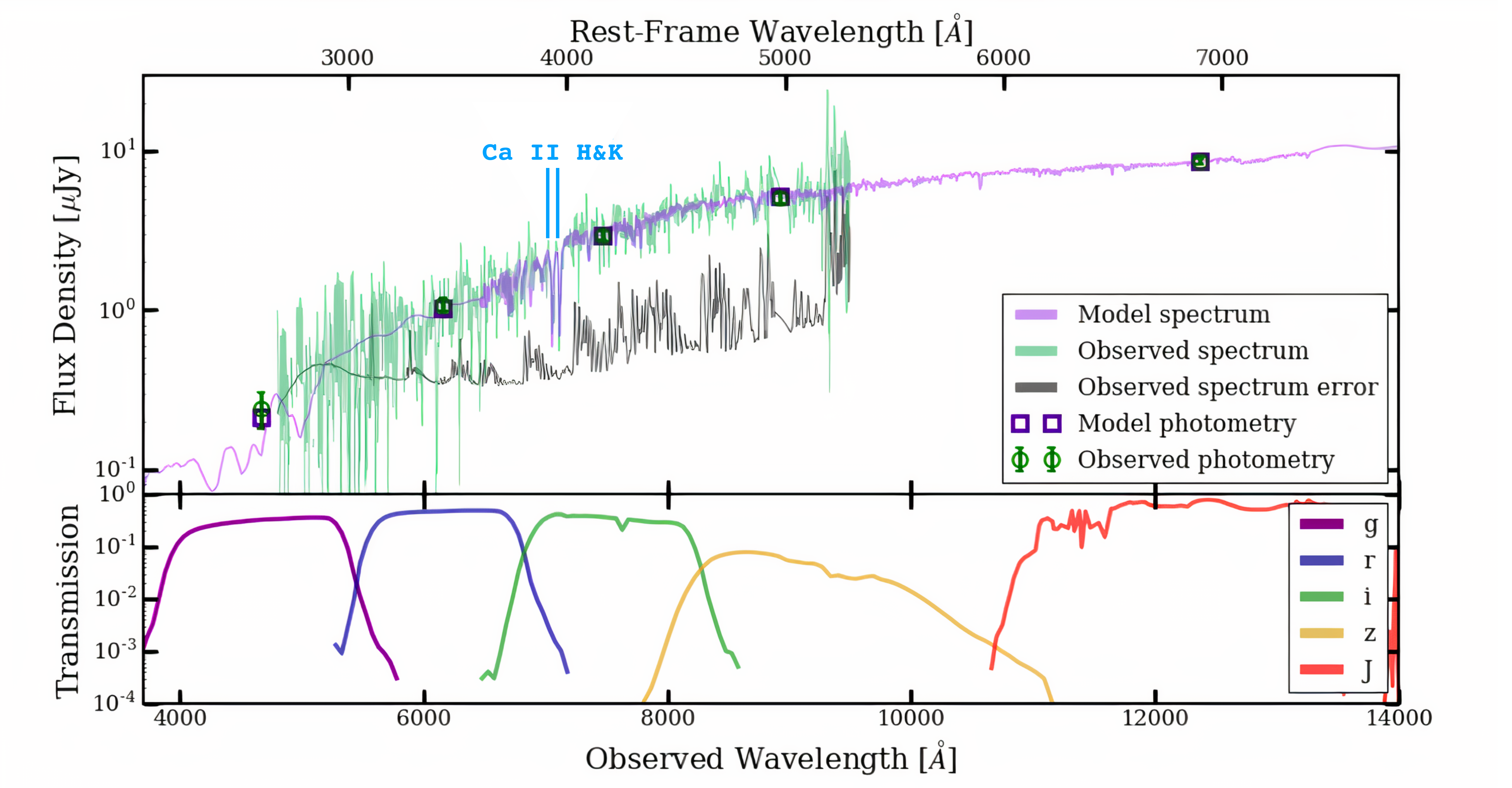}}
\vspace{-0.3in}
\caption{{\it Top:} The spectrum and $grizJ$ band photometry of the putative host galaxy of GRB\,161104A, G1 (green line and data points), error spectrum (black line) and the best-fit stellar population model spectrum and photometry from {\tt Prospector} (purple line and squares), characterized by median values of the posterior distributions shown in Figure~\ref{G1_corner} at a fixed redshift of $z=0.793$. The spectrum exhibits clear Ca II H and K absorption lines, marked by the blue lines, and a 4000 \AA\ break at an observed wavelength range of $\sim 7000 - 7200$ \AA. There is excellent consistency in the observed photometric colors, spectral continuum and features, and {\tt Prospector} model. The model and observed spectra are smoothed with a Savitzky-Golay filter with a bin size of 11 for clarity, although fits were performed on the unbinned data. {\it Bottom:} SDSS $griz$ band and Fourstar $J$ band filter response curves that are used in the fit.
}
\label{G1_SED}
\end{figure*}

\subsection{Stellar Population Properties}
\label{sec:fits}
Here, we describe fits to all available data for G1-G5. For all fits we use Sloan Digital Sky Survey (SDSS) filter response curves \citep{SDSSFilters}, which \texttt{Prospector} uses to calculate fluxes from the model spectrum and determine the effective wavelengths for the photometry in the $griz$ bands. We also use the relevant Fourstar response curve for $J$-band photometry \citep{FourstarFilters}.

\begin{figure*}
\centering
\includegraphics[width=0.8\textwidth]{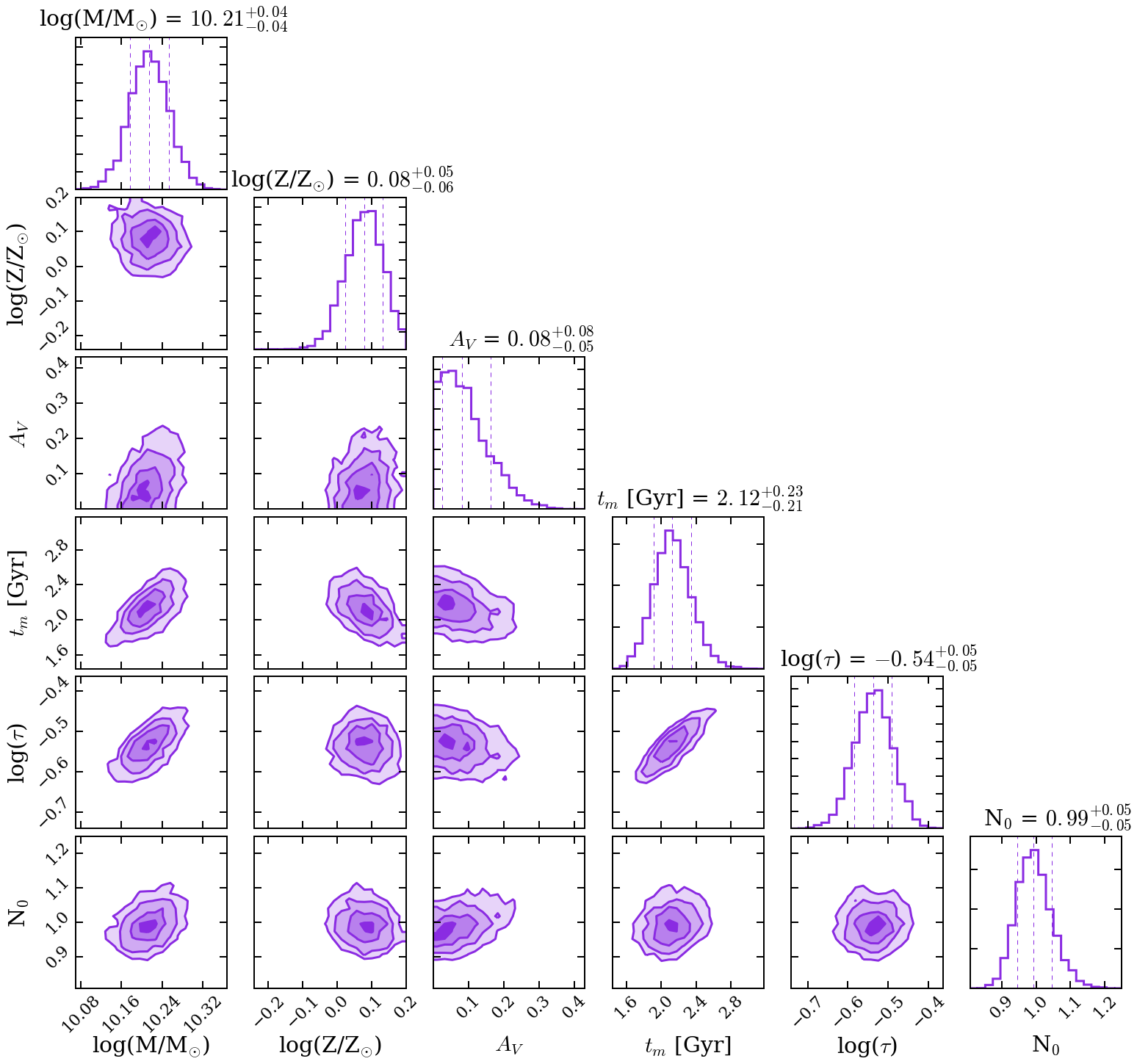}
\caption{\texttt{Prospector} results and parametric correlations from the joint fit of photometric and spectroscopic data of the putative host of \grb, G1. The posterior distributions of mass given in $\log(M/M_\odot)$, metallicity in  $\log(Z/Z_\odot)$, dust extinction $A_V$ in mag, mass-weighted age $t_m$ in Gyr, SFH parameter $\tau$, and spectral normalization factor $N_0$ are shown with the first 3000 out of 10,000 iterations removed. The contours correspond to the $1\sigma$,  $2\sigma$, and $3\sigma$ confidence levels, going from the darkest to lightest shades. The values at the top of each distribution represent the median value of the distribution and $1\sigma$ uncertainties, which are also shown as the dashed lines in each distribution plot. These values are typical for quiescent galaxies.}
\label{G1_corner}
\end{figure*}

Since G1 has a moderate signal-to-noise ratio (S/N) spectrum with clear absorption lines, we use both the spectrum and photometry in the fitting. We use a $10^{\text{th}}$-order Chebyshev polynomial to fit the spectral continuum, as this will capture fluctuations in the spectrum on the scale of $\sim 500$ \AA. We present a comparison of the model and observed spectroscopy and the model and observed photometry in Figure~\ref{G1_SED}, where the model plotted is characterized by the median of parameter posteriors (Table~\ref{tab:prospectres}). The median model exhibits remarkable consistency with the observed data. The posterior distributions of the parameters from the nested sampling fitting and the parameter correlations are shown in Figure~\ref{G1_corner}, with median and $1\sigma$ uncertainties denoted. 

Our best-fit solution shows that G1 has an older stellar population, with a median age of $2.12^{+0.23}_{-0.21}$ Gyr and a stellar mass of $\approx 1.62\times10^{10} M_\odot$, typical for short GRB hosts \citep{Berger_2014}. The metallicity of log($Z/Z_{\odot}) \approx 0.08$ is consistent with the $M-Z$ relation at $z \approx 0.7$ given the stellar mass and redshift of G1 \citep{MZRelation, MetModel}. The low dust extinction of $A_V \approx 0.08$~mag is expected, as quiescent galaxies do not contain much dust.  We also find $\log(\tau) \approx -0.54$. Using the SFH, we determine an SFR at $z=0.793$ of $\approx 0.099\,M_{\odot}$~yr$^{-1}$. We also derive a $3\sigma$ upper limit on the SFR from the absence of [OII]$\lambda 3727$, by calculating the expected integrated flux based on the error spectrum within a $10$\AA\ wavelength region of the [OII] doublet. Using the relationship between [OII] luminosity and SFR \citep{ken98}, we obtain SFR$\lesssim 0.4\,M_{\odot}$~yr$^{-1}$, which is in agreement with the results from the spectral energy distribution (SED) fitting.

\begingroup
\setlength{\tabcolsep}{6pt} 
\renewcommand{\arraystretch}{1.5} 
\begin{deluxetable*}{l|cccccc}[!t]
\tabletypesize{\normalsize}
\tablecolumns{7}
\tablewidth{0pc}
\tablecaption{GRB 161104A Surrounding Galaxies Stellar Population Properties
\label{tab:prospectres}}
\tablehead{
\colhead {Galaxy}	 &
\colhead {z}	 &
\colhead{$t_m$ [Gyr]} &
\colhead {$\log(Z/Z_{\odot})$}  &
\colhead{$\log(M/M_{\odot})$} &
\colhead {$\log(\tau)$}		    &
\colhead {$A_V$}
}
\startdata
G1$^\star$ & $0.793 \pm 0.003^\dagger$ & $2.12^{+0.23}_{-0.21}$ & $0.08^{+0.05}_{-0.06}$ & $10.21^{+0.04}_{-0.04}$ & $-0.54^{+0.05}_{-0.05}$ & $0.08^{+0.08}_{-0.05}$  \\ 
G2 & $0.79^{\dagger\dagger} $ & $0.35^{+0.78}_{-0.26}$ & $-0.58^{+0.32}_{-0.28}$ & $8.50^{+0.33}_{-0.40}$ & $0.24^{+0.31}_{-0.46}$ & $0.72^{+0.43}_{-0.41}$ \\ 
G3 & $0.79^{\dagger\dagger}$ & $1.96^{+0.94}_{-0.98}$ & $0.08^{+0.08}_{-0.14}$ & $10.22^{+0.12}_{-0.16}$ & $0.39^{+0.36}_{-0.27}$ & $1.83^{+0.30}_{-0.35}$ \\ 
G4 & $1.623^{\dagger\dagger}$ & $0.73^{+0.23}_{-0.13}$ & $-1.74^{+0.29}_{-0.19}$ & $9.98^{+0.09}_{-0.08}$ & $-0.85^{+0.20}_{-0.11}$ & $0.07^{+0.12}_{-0.06}$ \\ 
G5 & $0.788 \pm 0.003^\dagger$ & $2.09^{+0.30}_{-0.12}$ & $0^{\dagger\dagger}$ & $11.15^{+0.06}_{-0.05}$ & $-0.77^{+0.27}_{-0.17}$ & $0.92^{+0.10}_{-0.09}$ \\ 
\enddata
\tablecomments{We present the median values of the posterior distributions and their $1\sigma$ uncertainties from our \texttt{Prospector} runs. We use the spectroscopically determined redshift for G1 and G5. For G2 and G3, we use the inferred redshift based on the color of the galaxy. We use the potential redshifts based on the one emission line found for G4. Other parameter values are fixed only if the data quality is not sufficient enough to run a full six-parameter fit.\\
$^\star$ \footnotesize{We consider G1 to be the most likely host galaxy of \grb.} \\
$^\dagger$ \footnotesize{Spectroscopically determined redshift.} \\
$^{\dagger\dagger}$ \footnotesize{Fixed value in \texttt{Prospector} based on inference.}}
\end{deluxetable*}
\endgroup

Since G2, G3, and G4 lie on the outskirts of the XRT position (Figure \ref{composite}), we also determine the stellar population properties with all the available data. For G2, we only use the photometry in the fitting, as the spectroscopy is too low S/N to extract. Furthermore, because we do not have a redshift for this galaxy, we perform a fit where we allow the redshift to be a free parameter and another where we set it to $z=0.79$, as it has similar colors to the other galaxies in the field and is likely part of the galaxy cluster (see Section \ref{sec:cluster}). For the fit in which redshift is free, we find that the median and $1\sigma$ uncertainty is $z=0.90^{+0.29}_{-0.41}$, with a noticeable peak in the posterior distribution at $z \approx 0.79$. For the fit with fixed $z=0.79$, we determine the mass-weighted age to be $t_m = 0.35^{+0.78}_{-0.26}$~Gyr, younger than G1, with a lower inferred mass of $M \approx 3.16 \times 10^{8} M_{\odot}$ (Table~\ref{tab:prospectres}).

For G3, the average S/N of the LDSS3 spectrum is similarly low, $\approx 1.07$, and there are no definitive spectral features other than a reddening that could be interpreted as a 4000 \AA\ break at $\approx 7100$ \AA; thus, we only use the photometry in the fit. Similar to G2, we initially perform a fit in which redshift is a free parameter, finding $z = 0.55^{+0.07}_{-0.08}$, with some probability out to $3\sigma$ that $z=0.79$. Motivated by the possible 4000 \AA\ break, as well as the presence of the galaxy cluster (Section \ref{sec:cluster}), we set $z=0.79$ as a fixed parameter, finding $t_m = 1.96^{+0.94}_{-0.98}$~Gyr and an inferred stellar mass of $M\approx 1.65 \times 10^{10} M_\odot$ (Table~\ref{tab:prospectres}). We also note that the choice of redshift between $z\sim 0.5$ and $0.8$ does not have a large effect on the remaining parameters. 

The IMACS and LDSS3 spectra of G4 exhibit only one clear emission line at 9776 \AA. We examine possible identifications for this line, among H$\alpha$ $\lambda6563$, [OII] $\lambda3727$, [OIII] $\lambda5007$, or H$\beta$ $\lambda4861$. If the line is [OIII] or H$\beta$, the locations of [OII] or H$\alpha$ at the corresponding redshifts would be covered by the spectra, yet are not detected. Given that the line strengths of [OII] or H$\alpha$ are typically stronger than H$\beta$ and [OIII], their absence rules out H$\beta$ and [OIII] as viable candidates for the observed line. If this line is [OII] or H$\alpha$, this gives redshifts of $z=1.623$ and $z=0.489$, respectively. For $z=1.623$, the locations of H$\alpha$, H$\beta$, and [OIII] are not covered by our spectrum. However, if $z=0.489$, the locations of H$\beta$, [OII], and [OIII] are covered but not detected. Thus, the most likely redshift is $z=1.623$ and we only perform a fit at this redshift. We find a young ($\approx 1$ Gyr) stellar population with a mass $M \approx 9.55 \times 10^{9} M_{\odot}$.

Similar to G1, we use the spectrum, photometry, and spectroscopically determined redshifts to fit for the stellar population properties of G5. We determine the uncertainty on the redshift in the same manner as we did for G1, finding $z = 0.788 \pm 0.003$. We set $Z = Z_\odot$ and use a $12^{\text{th}}$-order Chebyshev polynomial to fit the continuum of the spectrum. Our \texttt{Prospector} results show that G5 is about the same age as G1; however, it is much more massive, with $M \approx 1.41 \times 10^{11} M_\odot$ (Table~\ref{tab:prospectres}). The model fits both the observed spectrum and photometry well and correctly identifies the location and strength of the Ca II H and K absorption lines, implying an overall good fit. We also tested subsolar metallicities ($Z < Z_\odot$), which only increased the stellar population age. 

In summary, we find that the most probable host galaxy G1 is an early-type galaxy that is old ($\sim$ 2.12 Gyr), is massive ($\sim 1.6\times 10^{10} M_\odot$), and has a low amount of ongoing star formation ($\sim 0.099~M_{\odot}$~yr$^{-1}$) with stellar population properties that agree with those of quiescent, early-type galaxies. We also find that G2 and G3 are consistent with originating from a similar redshift as G1 and G5 ($z=0.79$), although the quality of the data does not allow us to make definitive conclusions. The placement of G3 in the color-magnitude diagram (Figure \ref{fig:CMD}; see Section \ref{sec:cluster}) further indicates that it is most likely at a similar redshift to that of G1 and G5. Finally, we find that G4 is likely a high-redshift, background galaxy at $z=1.623$.

\subsection{Literature Sample}
\label{sec:LB10}
\begin{figure}
\includegraphics[width=0.475\textwidth]{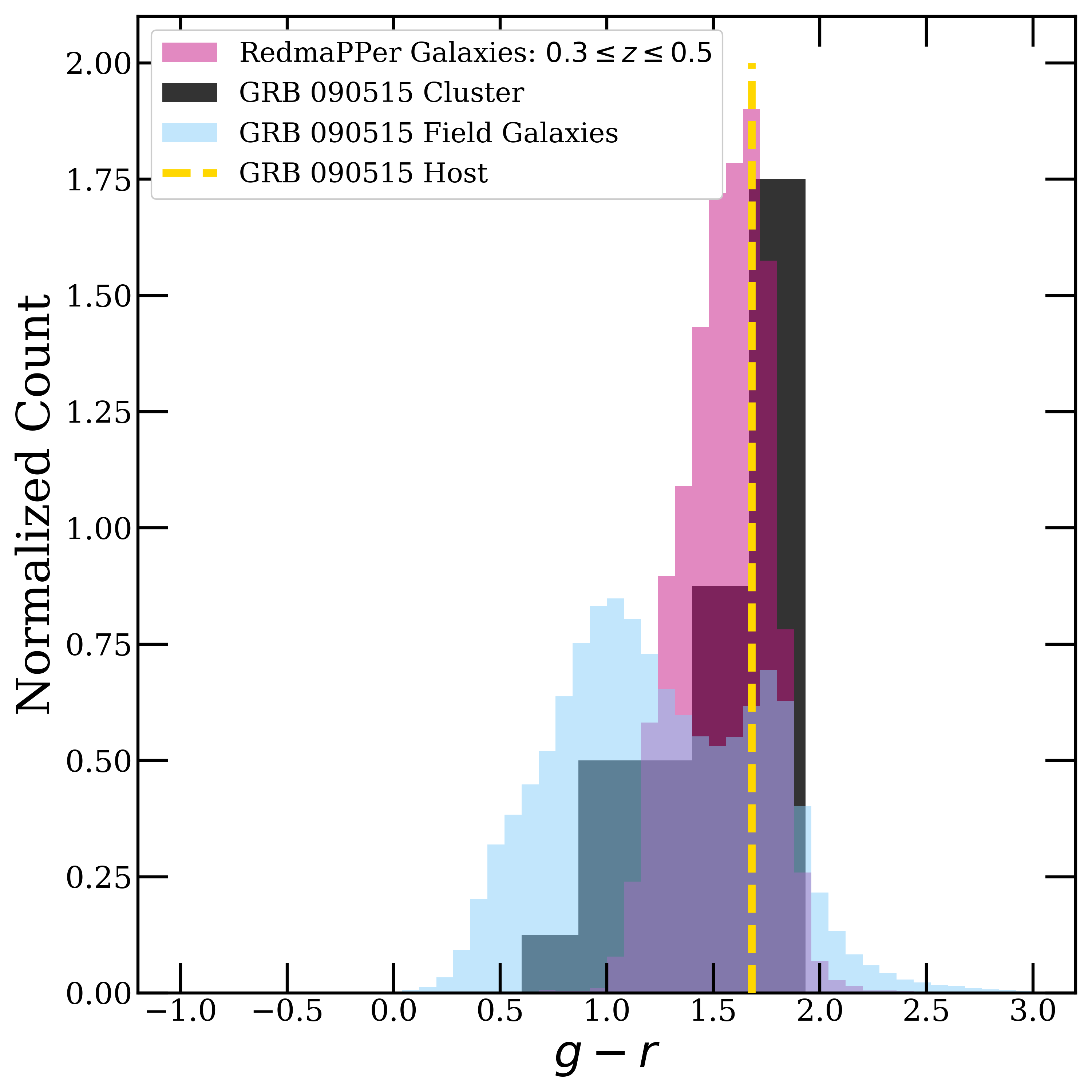}
\caption{Comparison of the putative cluster of GRB 090515 (black) to the redMaPPer cluster galaxies with $0.3 \leq z_{\text{phot}} \leq 0.5$ (pink) and the field galaxies surrounding the GRB (blue). The $g$-$r$ color of the host is marked by the yellow dashed line. We see a clear red sequence in the putative cluster. Furthermore, the colors of the putative cluster galaxies differ from those of the field galaxies, which are bluer, and are very similar to the redMaPPer galaxy clusters at contemporary redshifts.}
\label{fig:color_hist_090515}
\end{figure}

In order to compare the stellar population properties of the putative host G1 to those of other known short GRB hosts, we additionally collect photometric data from the literature, using the sample of \citet{LB10}, as well as the host of GRB 050813, another galaxy potentially in a cluster (\citealt{Prochaska2006,GRB050813}; see Section \ref{sec:cluster}). This sample contains multiband optical and near-IR (NIR) imaging for 19 hosts. We use this data set because it represents a uniformly analyzed photometric sample and represents the overall properties of the short GRB host population to provide an adequate comparison set.

\begin{figure*}
\makebox[\textwidth][c]{\includegraphics[width=0.95\textwidth]{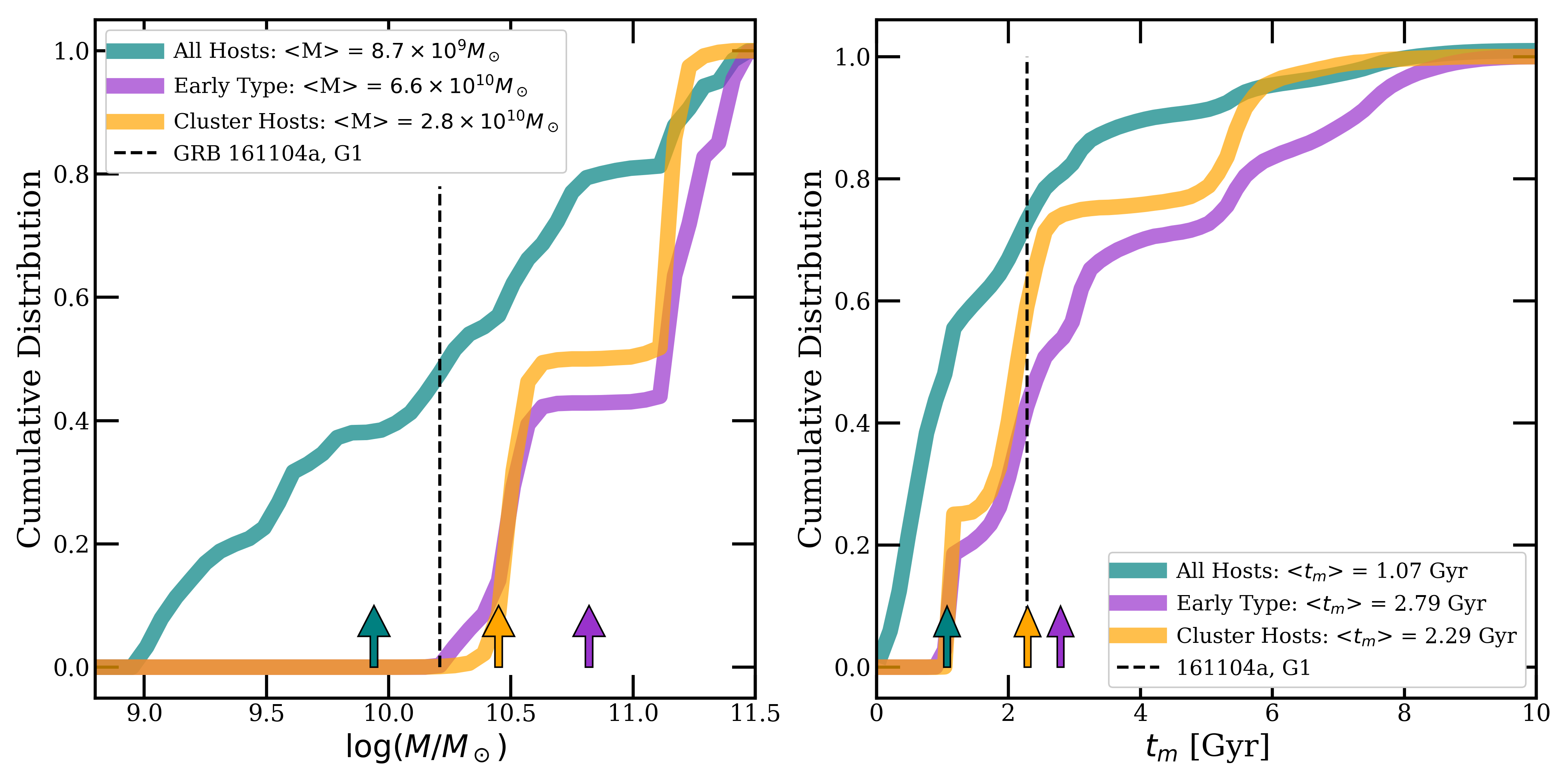}}
\caption{Cumulative distribution of mass, $\log(M/M_\odot)$, (left) and mass-weighted age, $t_m$, (right) of 21 short GRB host galaxies, which includes the putative host of \grb, G1.  Photometry for the fits were collected from \citet{LB10}, \citet{GRB050813}, and \citet{Prochaska2006}. Distributions including all hosts are green, for early-type galaxies are purple, and for host galaxies in clusters are orange. The arrows represent the median of each distribution. The dashed black lines show the median values of the host galaxy of GRB\,161104A, G1. }
\label{fig:cdf}
\end{figure*}

\begingroup
\setlength{\tabcolsep}{6pt} 
\renewcommand{\arraystretch}{1.3} 
\begin{deluxetable*}{l|ccccccc}[!t]
\tabletypesize{\normalsize}
\tablecolumns{7}
\tablewidth{0pc}
\tablecaption{Short GRB Host Galaxy Stellar Population Properties
\label{tab:prospect_LB10}}
\tablehead{
\colhead {GRB}	 &
\colhead {Type$^a$}	 &
\colhead {z}	 &
\colhead{$t_m$ [Gyr]} &
\colhead {$\log(Z/Z_{\odot})$}  &
\colhead{$\log(M/M_{\odot})$} &
\colhead {$\log(\tau)$}		    &
\colhead {$A_V$}
}
\startdata
050509B$^b$ & E & 0.225 & 1.16 & $0^{d}$ & 10.88 & -0.99 & 0.03 \\
050709 & L & 0.161 & 3.16 & $-0.60^{d}$ & 8.74 & 0.51 & 0.08\\
050724 & E & 0.257 & 7.58 & 0.16 & 11.09 & -0.57 & 0.04 \\
050813$^b$ & E & 0.716 & 2.30 & -0.06 & 10.22 & -0.39 & $0^{d}$ \\
051210 & ? & $1.34^{c}$ &  0.67 & -0.57 & 9.07 & 0.34 & $0^{d}$ \\
051221A & L &  0.546 & 0.76 & -1.92 & 9.32 & -0.96 & 0.11\\
060801 & L & 1.130 & 0.22 & -1.49 & 9.17 & 0.39 & 0.27 \\
061006 & L & 0.438 & 1.11 & $0^{d}$ & 8.84 & -0.03 & 0.37 \\
061210 & L & 0.410 & 0.58 & $0.60^{d}$ & 9.50 & -0.98 & 0.58\\
061217 & L & 0.827 & 0.30 & $-0.60^{d}$ & 8.94 & 0.11 & 0.06\\
070429B & L & 0.902 & 0.68 & $0^{d}$ & 10.49 & -0.67 & 1.08 \\
070714B & L & 0.923 & 0.88 & $0^{d}$ & 9.28 & 0.36 & $0.50^{d}$\\
070724 & L & 0.457 & 1.09 & 0.11 & 9.89 & 0.44 & 0.41\\
070729 & ? & $0.70^{c}$ & 2.21 & -0.34 & 10.13 & -0.77 & 0.53 \\
$070809^e$ & E & 0.473 & 3.14 & $0^{d}$ & 10.95 & -0.71 & $0^{d}$\\
071227 & L & 0.381 & 0.81 & 0.15 & 10.46 & -0.57 & 1.84 \\
080123 & L & 0.495 & 0.59 & -0.40 & 9.88 & -0.03 & 0.81 \\
090510 & L & 0.903 & 2.19 & 0.06 & 9.95 & 0.46 & $0.25^{d}$\\
$090515^{be}$ & E & 0.403 & 5.49 & $0^{d}$ & 10.87 & -0.58 & $0^{d}$\\
100117 & E & 0.920 & 2.23 & $0^{d}$ & 10.15 & -0.30 & $0^{d}$ \\
161104A$^b$ & E & 0.793 & 2.12 & 0.08 & 10.21 & -0.54 & 0.08 \\
\hline 
All Hosts &  & & 1.07 & 0.0 & 9.94 & -0.3 & 0.11 \\
Early-type & &  & 2.79 & 0.0 & 10.82 & -0.54 &  0.0 \\
Cluster & &  & 2.29 & 0.0 & 10.45 & -0.56 & 0.02 \\
\enddata
\tablecomments{Here we show the median stellar population property values determined through \texttt{Prospector} for the 20 sampled short GRB hosts described in \citet{LB10} and \citet{GRB050813}. Redshift values are fixed and all other parameters are free unless specified otherwise. We also show the median values for all hosts, quiescent/early-type, and cluster short GRB hosts. \\
$^a$ \footnotesize{Type classified in \citet{LB10} and \citet{Prochaska2006}.} \\
$^b$ \footnotesize{Galaxies in clusters.} \\
$^c$ \footnotesize{Redshift determined through \texttt{Prospector}.} \\
$^d$ \footnotesize{Fixed value in \texttt{Prospector}}.\\
$^e$ \footnotesize{From \cite{Zevin2019}.}}
\end{deluxetable*}

\begin{figure*}
\makebox[\textwidth][c]{\includegraphics[width=0.95\textwidth]{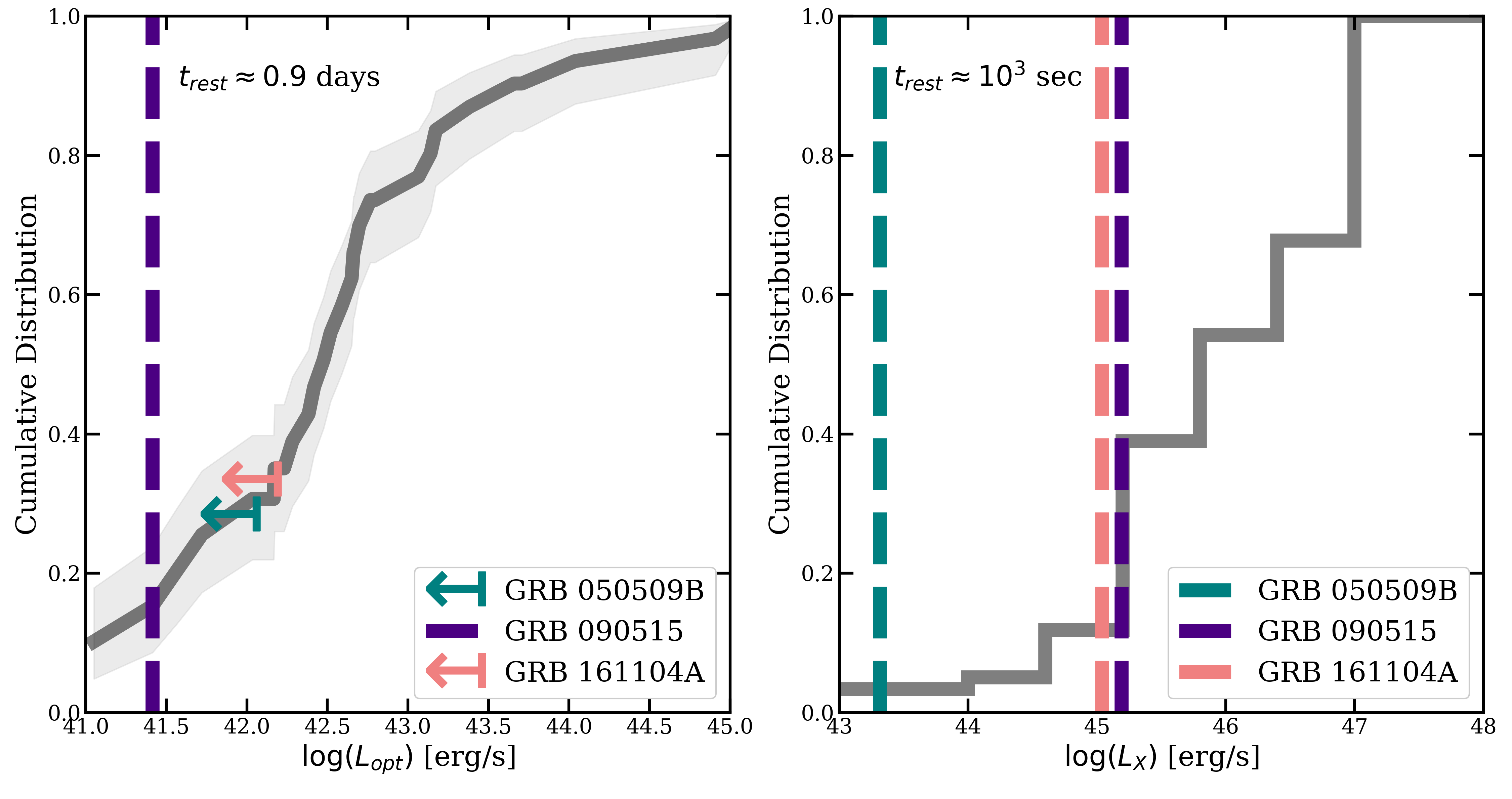}}
\caption{Cumulative distributions of short GRB optical afterglow luminosities, $L_\text{opt}$, at a fixed rest-frame time of $t_\text{rest} \approx 0.9$~days (left) and X-ray afterglow luminosities, $L_\text{X}$, at $t_\text{rest} \approx 10^3$~s (right) in dark grey. The fixed times are chosen to encompass as many cluster short GRBs as possible. We use survival statistics to account for upper limits in the optical afterglows taken at a similar value of $t_\text{rest}$, and therefore we also show the 68$\%$ confidence interval in light grey. We highlight the cluster GRBs with available data at those times: GRB\,050509B, GRB\,090515 and GRB\,161104A. Leftward-pointing arrows denote upper limits on afterglow emission. We find that for the optical and X-ray luminosities, the cluster GRBs are in the bottom $\approx 30\%$ and $\approx 11\%$ of the short GRB population, respectively. This is commensurate with older stellar populations and significant offsets from their hosts, which both translate to lower circumburst densities.} 
\label{fig:optical_lum}
\end{figure*}

We determine the stellar population properties of the 20 host galaxies with \texttt{Prospector}, using the same methods as detailed in Section \ref{sec:prospector}. For the 19 hosts in \citet{LB10}, we use the photometric observations listed in \citet{LB10}, \citet{Fong2010}, and \citet{Fong2013}, correcting for Galactic extinction when relevant \citep{sf11}. Each GRB host has at least four bands used in the fits. We use standard filter transmission curves from SDSS, the 2MASS \citep{2Mass}, and Wide Field Camera 3 (WFC3/IR and WFC3/UVIS; \citealt{WFC3}) in the \texttt{Prospector} fits of these galaxies. We include redshift as a fixed value in the \texttt{Prospector} fitting for galaxies with a spectroscopic redshift. We determine photometric redshifts for GRB 051210 and GRB 070729 with \texttt{Prospector}. For GRB 050813, we collect available photometry for the most likely host galaxy at a fixed redshift of $z=0.72$ \citep{GRB050813}. We find photometric observations of this galaxy in standard SDSS $grz$ and Wide-field Infrared Survey Explorer (WISE) $W1$ \citep{WISE} filters in the Legacy Survey Data Release 8 \citep{LegacySurveyDr8}. For this work, we concentrate on comparing the mass-weighted ages and stellar masses, and the median values are listed in Table {\ref{tab:prospect_LB10}}.

We next divide the 20 host galaxies into three groups to compare to the host galaxy of \grb\ (G1): all host galaxies, early-type host galaxies, and host galaxies in clusters. By default, there is overlap between the populations. Early-type classifications are defined as galaxies with limits on the SFRs of $\lesssim 0.1-1\,M_{\odot}$~yr$^{-1}$ \citep{LB10,Prochaska2006} and are marked in Table {\ref{tab:prospect_LB10}}. Within the entire sample, there are three hosts in addition to G1 that have been identified as cluster members: those of GRB 050509b, GRB 050813, and GRB 090515 \citep{Prochaska2006, Berger2007, GRB050813}. We confirm the cluster association of GRB\,090515 in this work, applying the same methods as described in Section \ref{sec:cluster} for \grb. From the Dark Energy Spectroscopic Instrument (DESI) imaging dataset \citep{DESI}, we find all galaxies to be within a $\approx 178''$ radius around the brightest galaxy near the GRB location, where $0.3 \leq z_{\text{phot}} \leq 0.5$ (assuming the photometric redshifts from \citealt{zhou2020clustering}). There are $\approx 50$ galaxies in this sample. We compare this sample to the $g$-$r$ colors of the redMaPPer cluster galaxies within the same photometric redshift range and the field galaxies around GRB\,090515. We find that the putative cluster has a clear red sequence that matches the colors of the redMapper cluster galaxies well and differs from the field galaxies, which we demonstrate in Figure \ref{fig:color_hist_090515}. We further quantify GRB\,090515's association to the cluster using the $P_{cc}$ method described in Section \ref{sec:clustermem}, finding that $P_{cc} = 2.4 \times 10^{-3}$, which implies low chance coincidence. Thus, we confirm both the presence of a cluster around GRB\,090515 and the association to the cluster.

We construct cumulative distributions for each of the three groups by combining the individual host posterior distributions and normalizing the areas under the probability distributions to unity. The resulting cumulative distributions for stellar mass and mass-weighted age for each category, along with the comparison to the values of G1 are shown in Figure~\ref{fig:cdf}. We find median values and 68\% confidence intervals for the ``all hosts'' population, $\log(M/M_\odot) = 9.94^{+0.88}_{-0.98}$ and $t_m = 1.07^{+1.98}_{-0.67}$~Gyr; for the early-type population, $\log(M/M_\odot) = 10.82^{+0.17}_{-0.67}$ and $t_m = 2.79^{+3.57}_{-1.37}$~Gyr; and for the cluster population, $\log(M/M_\odot) = 10.45^{+0.40}_{-0.29}$ and $t_m = 2.29^{+3.19}_{-1.01}$~Gyr. Overall, the host galaxy of \grb\ is a younger and less massive galaxy than the medians of the samples of both early-type and cluster galaxies (Figure~\ref{fig:cdf}).

\section{Discussion}
\label{sec:discussion}

\subsection{Comparison of \grb\ to Cluster Short GRBs}

Given that the XRT position of \grb\ fully encompasses a galaxy at $z=0.793$, coupled with the spatial coincidence with a galaxy cluster at $z\approx 0.8$, we find this to be the most likely redshift and host association of \grb. We note that we cannot rule out an origin from G4 (tentatively at $z=1.623$) based on chance probability arguments alone. However, the likelihood of association to G4 is comparatively small given the decreased sensitivity of {\it Swift} in detecting $z>1.5$ short GRBs \citep{brf14}; indeed, there are only three short GRBs to date confirmed to be at $z>1.5$ \citep{Paterson2020}. At $z=0.793$, the {\it Swift}/XRT afterglow luminosity is $L_X\approx 1.1 \times 10^{45}$~erg~s$^{-1}$ (0.3-10~keV; \citealt{ebp+07}) at a rest-frame time of $\approx 0.2$~hr after the burst, which is in the lower $\approx 8\%$ when compared to short GRB afterglows with known redshifts at similar rest-frame times. The deep limit on optical afterglow emission of $r>25.4$~mag (see Section \ref{sec:photometry}) also constrains the optical emission to be similarly faint, with $L_{\rm opt} < 1.46\times10^{42}$~erg~s$^{-1}$ at a rest-frame time of $0.72$~days post-burst. This places \grb\ in the bottom $\approx 30\%$ for both detected afterglows and upper limits at the same time, as shown in Figure \ref{fig:optical_lum}. These results can naturally be explained by an event originating from an elliptical galaxy, which on average have lower gas densities than star-forming galaxies, which results in fainter afterglow emission.

We next compare \grb\ to the $\gamma$-ray, afterglow and host galaxy properties of short GRBs known to be associated with galaxy clusters. With the addition of both GRB\,090515 and \grb, there are five such {\it Swift} short GRBs: GRB 050509B ($z=0.225$; \citealt{Bloom2006}), GRB 050813 ($z=0.72$; \citealt{GRB050813}), GRB 050911 ($z=0.1646$; \citealt{pkl+06}), GRB 090515 ($z=0.403$; this work), and GRB 161104A ($z=0.793$). GRB\,050911 is classified as a short GRB with extended emission \citep{Lien2016}, and the BAT $\gamma$-ray position is coincident with a low-redshift cluster at $z=0.1646$ with a confidence of association at the $\sim 3\sigma$ level \citep{Berger2007}. While the positional uncertainty precludes a clear host association, we still include this event as a cluster short GRB in our subsequent comparisons. We find that cluster short GRBs have $T_{90} \leq 0.450$~s, except for GRB\,050911, which has $T_{90} \approx 16$~s, with a $1.5$~s initial spike. The fraction of {\it Swift} short GRBs that have $T_{90}<0.450$~sec is 53.7\%. Thus, cluster short GRBs appear to trace the bottom half of the $T_{90}$ distribution, solidifying their membership as true short GRBs. 

Turning to their afterglow properties, with the exception of GRB\,050911, all of the cluster short GRBs have detected XRT afterglows, with localizations of roughly a few arcsec. However, their X-ray afterglows are uniformly faint with $L_{\rm X} = 2.1 \times 10^{43}$~erg~s$^{-1}$-$1.1 \times 10^{45}$~erg~s$^{-1}$, far below the median values at contemporary times after the bursts lying at $\leq 11\%$ (Figure \ref{fig:optical_lum}). Similarly, there are deep limits on optical afterglow emission for GRB\,050509B \citep{Bloom2006} and GRB\,161104A, while GRB\,090515 had a detected optical afterglow \citep{Rowlinson2010} but is among the least luminous afterglows compared to short GRBs. To adequately compare them to the short GRB sample, we collect all optical afterglow detections with $t_{\rm rest} \approx 0.9$~days (chosen to encompass as many cluster short GRBs as possible), as well as upper limits. We use the Kaplan-Meier estimator to create an inverse survival function to properly incorporate both detections and upper limits (Figure \ref{fig:optical_lum}). Compared to this function, cluster short GRBs are in the bottom $\approx 30\%$ of the population, while GRB\,090515 is at $11\%$. These faint or undetected afterglows are likely a direct reflection of their low environmental densities (although we note that only GRB\,050509B has an inferred density constraint of $n<0.015$~cm$^{-3}$; \citealt{smf+20}), and commensurate with their significant projected offsets from their respective host galaxies. While this value is highly uncertain for \grb\ ($1.66\pm16.66$~kpc), the offsets extend to $\approx 37-75$~kpc for the remaining events in clusters, significant compared to the median short GRB offset of $\approx 6$~kpc \citep{Fong2013}. These large offsets of cluster short GRBs could be explained by an observational bias, as galaxy clusters are composed of more massive galaxies on average, and thus the association to a nearby, massive cluster galaxy at a large offset will be more likely than a fainter, field galaxy at a smaller offset for a given short GRB. On the other hand, this may also indicate a specific physical mechanism within galaxy clusters that favors short GRBs, or a product of the fact that cluster galaxies have older stellar populations and thus long delay times that would naturally explain large offsets.

Of the cluster short GRBs, three have robust host associations: GRB\,050509B, ($P_{cc}=5\times10^{-3}$; \citealt{Bloom2006}, GRB\,090515 ($P_{cc}=0.05$; \citealt{Fong2013}), and now \grb\ with $P_{cc}=0.06$, while GRB\,050813 has a tentative host association based on the most likely host in its vicinity ($P_{cc}=0.20$; \citealt{Prochaska2006,GRB050813}). Given that the frequency of elliptical galaxies is higher by a factor of two in clusters than in the field \citep{Dressler1980, Whitmore1993}, it is perhaps unsurprising that {\it all} of the cluster short GRB hosts discovered to date are early-type galaxies with little or no signs of ongoing star formation. Their redshifts span $z\approx 0.15-0.79$, with \grb\ setting the high-redshift end of this range.

Cluster short GRB hosts also appear to follow the morphology-density relation, in which dense cluster centers contain a larger fraction of massive, elliptical galaxies, compared to an increasing fraction of star-forming galaxies toward less dense regions \citep{Dressler1980}. The host galaxy of GRB\,050509B has $\approx 5L^*$ (where $L^*$ is the characteristic luminosity in the galaxy luminosity function; \citealt{Berger_2014}) and a stellar mass of $\approx 7.6 \times 10^{10}\,M_{\odot}$. Given that it is significantly brighter and more massive than its surrounding galaxies, it is likely the BCG of its cluster \citep{Bloom2006}. The host of GRB\,090515 may also be a BCG, with a similarly large stellar mass of $\approx 7.4 \times 10^{10}\,M_{\odot}$, and a high spatial density of luminous galaxies in its vicinity. In contrast, the host galaxy of \grb\, G1, is at a considerable distance ($\approx 1$~Mpc) from its presumed BCG and has a measurable, low rate of star formation. Its stellar population is presumably less affected by the high densities at the center of the cluster and is also commensurate with its higher redshift, younger age, and lower stellar mass compared to those of other cluster short GRBs. 

\subsection{\grb\ in the Context of Short GRBs and Cluster Transients}

Overall, the short GRB host galaxy population is diverse, with short GRBs originating from star-forming galaxies with signatures of ongoing star formation (star-forming, late-type), as well as host galaxies with no sign of star formation (quiescent, early-type). This diversity is a reflection of a broad progenitor delay time distribution and can be naturally explained in the context of a BNS merger progenitor with a wide range of merger timescales \citep{bpb+06,Nakar2006, Berger2007_HighZ, Zheng2007,Fong2013}. As expected for early-type galaxies, the host of GRB\,161104A has a higher stellar mass and older mass-weighted age than the respective medians for the entire population (at the 63\% and 71\% level, respectively), but a lower average stellar mass and younger age when compared to short GRB early-type and cluster hosts (Figure~\ref{fig:cdf}). Thus, the stellar population properties of the host of \grb\ appear to be more in line with those of field galaxies than cluster hosts, also consistent with its location with respect to the BCG within its cluster.

With the addition of GRB\,161104A, there are now 10 early-type short GRB host galaxies (the others are GRB\,050509B, GRB\,050724A, GRB\,050813, GRB\,060502B, GRB\,070809, GRB\,090515, GRB\,100117A, GRB\,100625A, and GRB\,150101B; \citealt{Bloom2006,Berger_050724,Bloom2007,GRB050813,ber10,fbc+11,Fong2013_Dem,fmc+16}). In comparison with the 38 short GRBs with enough spectroscopic information to constrain their SFRs, the early-type fraction composes 26\%. Given that current state-of-the-art cluster catalogues are only complete to $z<0.6- 0.8$ \citep{Redmapper,DES2020}, compared to the short GRB redshift distribution, which extends to $z\approx 2$ \citep{Paterson2020}, it is probable that there are additional associations to clusters among the {\it Swift} short GRB population. Given that short GRBs trace the low-luminosity regime of afterglows, events associated with cluster hosts or ICM environments are more likely to have undetectable afterglows than short GRBs with spatially coincident host galaxies.

We can obtain a lower limit on the fraction of short GRBs associated with clusters, set by the five known associations among the population with redshifts of $\gtrsim 13\%$. A more conservative lower limit comes from a comparison to the {\it Swift} short GRB population, or those detected to date that otherwise have no observing constraints that would preclude optical follow-up. This population totals 99 events, giving a conservative lower limit on the fraction of cluster short GRBs of $\gtrsim 5\%$. Finally, a somewhat less meaningful upper limit can be set by taking into account studies that have ruled out cluster associations for four events \citep{Berger2007,fbc+11}, of $\lesssim 89\%$. We also find that cluster short GRB hosts compose $20\%$ of the stellar mass of all short GRB hosts within the sample described in Table \ref{tab:prospect_LB10}. Overall, these fractions align with the fraction of stellar mass in clusters of $\approx 10-20\%$, and indirectly constrain the minimum fraction of short GRBs with long delay times.

Although we have not taken into account the latest short GRBs in our sample, it is interesting to note that the gravitational wave BNS merger GW 170817 \citep{TheLIGOScientific:2017qsa}, associated with GRB 170817A and with emission over a broad range of wavelengths \citep{GBM:2017lvd}, was also found in an old, early-type, elliptical galaxy, NGC 4993,
with a stellar mass similar to G1 ($\sim 3.8 \times 10^{10} ~M_\odot$; \citealt{Blanchard_2017,palmese}).
NGC 4993 is part of a galaxy group containing 22 members \citep{Kourkchi_2017}, as opposed to $\sim 70$ for the galaxy cluster analyzed here. While it is likely that the host of GRB 170817A lives in a dark matter halo with significantly smaller mass than that of \grb, it is worth noting that in both cases the GRB can be associated with large-scale structure overdensities and possibly long delay times \citep{Blanchard_2017}.

In addition to short GRBs, a few other transients have been linked to galaxy clusters and signify the existence of subpopulations with long delay times. Moreover, their nucleosynthetic outputs have been linked to the chemical enrichment of the ICM. For instance, some Ca-strong transients are associated with galaxy cluster environments at large offsets from their host galaxies \citep{Lunnan2017} with no discernible signs of star formation at their explosion sites \citep{pgm+10}. While the progenitors of Ca-strong transients are unknown, the link of a substantial fraction to old stellar populations at large radial offsets has pointed to an origin from white dwarf progenitors in globular clusters \citep{Shen2019}. The isolated locations of these events, coupled with their rates, have also been shown to reconcile the Ca abundances in the ICM \citep{mkk14,Frohmaier2018}. Meanwhile, the rates of SNe Ia in clusters have also been used to reconcile observed Fe abundances in the ICM and have provided constraints on a `delayed' channel. Among early-type galaxy and S0 hosts, SNe Ia have an enhanced rate in clusters versus the field \citep{Sand2008} and also compared to other types of core-collapse SNe (Type Ib/c and Type II; \citealt{SNRateGC}). This is commensurate with their older, single- or double-degenerate white dwarf progenitors \citep{Maoz2014}. In particular, the evolution of SN Ia cluster rates with redshift has been used to explain the distribution of Fe in the ICM, as well as the SN Ia delay time distribution and respective predictions for single- versus double-degenerate progenitors \citep{sgm+10,baa+12,fm18}.

Looking forward, assuming that the majority of short GRBs are linked to BNS merger progenitors, the true fraction of short GRBs that occur in galaxy clusters may similarly play a role in $r$-process element enrichment of the ICM and intracluster light (ICL). In addition, the large physical offsets of cluster short GRBs from their most probable hosts, relative to those associated with field galaxies, is of interest in understanding their origins. A natural explanation for the occurrence of short GRBs at tens of kiloparsecs away from their host galaxies, but within galaxy clusters, are the large systemic velocities that BNS systems may experience \citep{Zevin2019}, coupled with the older stellar population in clusters that contribute to longer delay times. Galaxy cluster studies have also shown that the ICL and BCGs compose $\approx 30-50\%$ of the cluster stellar mass \citep{zyp+19}. In particular, the origins of the ICL and ICM has been linked to the disruption of dwarf galaxies \citep{Gregg1998} or the tidal stripping of outer material from more typical galaxy cluster members \citep{GallagherOstriker1972,DeMaio2015,DeMaio2018}. Since $30\%$ of short GRBs occur in the halos of their host galaxies \citep{Fong2013, Zevin2019}, beyond a few effective radii, the locations of these short GRBs are situated amid prime material for stripping to contribute to the ICL and ICM. 

\section{Conclusions}
\label{sec:conc}
We have presented identification of the early-type, quiescent host galaxy of \grb\ and a detailed investigation of the large-scale galaxy cluster environment surrounding \grb. We have also presented new modeling of 20 additional short GRB host galaxies, as well as the identification of the large-scale environment of GRB\,090515 as a galaxy cluster. We summarize with the following conclusions:

\begin{itemize}
    \item The putative host of \grb\ is an early-type, quiescent galaxy at $z = 0.793 \pm 0.003$. By jointly fitting the host photometry and spectroscopy from Magellan with {\tt Prospector}, we determine a stellar mass of $M \approx 1.6\times10^{10}~M_\odot$, a mass-weighted age of $t_m \approx 2.12$~Gyr, metallicity of $\log(Z/Z_\odot) \approx 0.08$, dust extinction $A_V \approx 0.08$, and ongoing SFR$\approx 0.099 M_\odot$~yr$^{-1}$.
    \item Using deep optical and NIR observations, we determine that the host of \grb\ and most of the galaxies surrounding the XRT position are at similar redshifts of $z \sim 0.8$. We confirm their membership in a cluster at a median redshift of $z=0.79$ using DES cluster catalogues. We also show that GRB\,090515 resides in a cluster at $z \sim 0.4$. This makes \grb\ the fifth short GRB associated with a cluster and the highest-redshift short GRB known to be associated with a galaxy cluster. 
    \item We model the broadband SEDs of 20 additional short GRB hosts and determine median stellar population properties of $\log(M/M_\odot) = 9.94^{+0.88}_{-0.98}$ and $t_m = 1.07^{+1.98}_{-0.67}$~Gyr. Compared to the other early-type and cluster host galaxies studied in this paper, the host of \grb\ is slightly younger and less massive. However, its mass and age are consistent with the median values of early-type, quiescent short GRB host galaxies.
    \item In comparison to short GRB optical and X-ray afterglow luminosities, the afterglows of cluster short GRBs are in the faintest $\approx 30 \%$ and $\approx 11\%$ of observed optical and X-ray luminosity distributions, respectively, consistent with the expectation that they have lower circumburst densities.
    \item We calculate a lower limit on the fraction of {\it Swift} short GRBs in galaxy clusters of $\approx 5-13\%$, and the fraction of stellar mass in cluster short GRB hosts of $\gtrsim 20\%$. This is consistent with the fraction of stellar mass in galaxy clusters at $0.1 \leq z \leq 0.8$ of $\approx 10-20\%$. 
\end{itemize}

The discovery of \grb\ in a galaxy cluster adds to the small number of short GRBs associated with galaxy clusters. Associating short GRBs with clusters is currently very difficult, given that few cluster catalogues reach the cosmological distances at which short GRBs occur and cover a wide enough region of the sky to capture the locations of the majority of short GRBs. For example, SDSS cluster catalogues are highly incomplete beyond $z \sim 0.4$ \citep{SDSS_Cluster}, and current DES catalogues, which already provide a cluster sample over $\sim1800$~deg$^2$, only reach $z \sim 0.6$ \citep{DES2020}, and are complete for cluster members down to stellar masses of $10^{10}~M_\odot$ \citep{2020MNRAS.493.4591P}. Moreover, confirmation via multislit spectroscopy or the detection of ICL in X-rays becomes extremely challenging beyond $z\gtrsim 0.5$, as spatial clustering is less apparent and contamination with foreground and background galaxies is an issue at high redshifts.

Looking forward, the final DES Year 6 cluster catalogue, will be ideal for cluster associations of short GRBs, as it will cover over $5000$~deg$^2$ out to $z\sim0.9$. Furthermore, the Vera C. Rubin Observatory (VRO; \citealt{LSST}) will identify clusters out to even larger redshifts over most of the southern hemisphere, and will thus be crucial to short GRB host and cluster associations. We therefore expect that short GRB rates in clusters will be significantly more constrained in the next decade, as more wide-field, deep surveys and catalogues are released. In parallel, modeling that incorporates both photometric and spectroscopic data of short GRB hosts is needed to fully understand their star formation histories, which are known to differ between galaxies in clusters and those in the field. Finally, given that BNS mergers are known sources of $r$-process element nucleosynthesis, the actual cluster rate will be crucial in understanding enrichment of the ICM, as well as the contribution of a substantial `delayed' channel of mergers.

\section*{Acknowledgments}
\noindent We thank the anonymous referee for valuable feedback and suggestions. We also thank S. Coughlin for helpful discussions that aided this work. A.E.N. acknowledges support from the Henry Luce Foundation through a Graduate Fellowship in Physics and Astronomy. The Fong Group at Northwestern acknowledges support by the National Science Foundation under grant Nos. AST-1814782 and AST-1909358. This work was supported by the Fermi National Accelerator Laboratory, managed and operated by Fermi Research Alliance, LLC under contract No. DE-AC02-07CH11359 with the U.S. Department of Energy.

\noindent This work made use of data supplied by the UK Swift Science Data Centre at the University of Leicester. 

\noindent This paper includes data gathered with the 6.5 m Magellan Telescopes located at Las Campanas Observatory, Chile. 
\noindent Based on observations obtained at the international Gemini Observatory (PI Troja; GS-2016B-Q-28), a program of NOIRLab, which is managed by the Association of Universities for Research in Astronomy (AURA) under a cooperative agreement with the National Science Foundation on behalf of the Gemini Observatory partnership: the National Science Foundation (United States), National Research Council (Canada), Agencia Nacional de Investigaci\'{o}n y Desarrollo (Chile), Ministerio de Ciencia, Tecnolog\'{i}a e Innovaci\'{o}n (Argentina), Minist\'{e}rio da Ci\^{e}ncia, Tecnologia, Inova\c{c}\~{o}es e Comunica\c{c}\~{o}es (Brazil), and Korea Astronomy and Space Science Institute (Republic of Korea).

\noindent This project used public archival data from the Dark Energy Survey (DES). Funding for the DES Projects has been provided by the U.S. Department of Energy, the U.S. National Science Foundation, the Ministry of Science and Education of Spain, the Science and Technology Facilities Council of the United Kingdom, the Higher Education Funding Council for England, the National Center for Supercomputing Applications at the University of Illinois at Urbana–Champaign, the Kavli Institute of Cosmological Physics at the University of Chicago, the Center for Cosmology and Astro-Particle Physics at the Ohio State University, the Mitchell Institute for Fundamental Physics and Astronomy at Texas A$\&$M University, Financiadora de Estudos e Projetos, Fundação Carlos Chagas Filho de Amparo à Pesquisa do Estado do Rio de Janeiro, Conselho Nacional de Desenvolvimento Científico e Tecnológico and the Ministério da Ciência, Tecnologia e Inovação, the Deutsche Forschungsgemeinschaft and the Collaborating Institutions in the Dark Energy Survey.

\noindent The Collaborating Institutions are Argonne National Laboratory, the University of California at Santa Cruz, the University of Cambridge, Centro de Investigaciones Enérgeticas, Medioambientales y Tecnol\`{o}gicas–Madrid, the University of Chicago, University College London, the DES-Brazil Consortium, the University of Edinburgh, the Eidgen\"{o}ssische Technische Hochschule (ETH) Z\"{u}rich, Fermi National Accelerator Laboratory, the University of Illinois at Urbana-Champaign, the Institut de Ci\`{e}ncies de l'Espai (IEEC/CSIC), the Institut de F\'{i}sica d'Altes Energies, Lawrence Berkeley National Laboratory, the Ludwig-Maximilians Universit\"{a}t München and the associated Excellence Cluster Universe, the University of Michigan, the National Optical Astronomy Observatory, the University of Nottingham, The Ohio State University, the OzDES Membership Consortium, the University of Pennsylvania, the University of Portsmouth, SLAC National Accelerator Laboratory, Stanford University, the University of Sussex, and Texas A$\&$M University.

\noindent Based in part on observations at Cerro Tololo Inter-American Observatory, National Optical Astronomy Observatory, which is operated by the Association of Universities for Research in Astronomy (AURA) under a cooperative agreement with the National Science Foundation.

\noindent The Legacy Surveys consist of three individual and complementary projects: the Dark Energy Camera Legacy Survey (DECaLS; NSF's OIR Lab Proposal ID No. 2014B-0404; PIs: David Schlegel and Arjun Dey), the Beijing-Arizona Sky Survey (BASS; NSF's OIR Lab Proposal ID No. 2015A-0801; PIs: Zhou Xu and Xiaohui Fan), and the Mayall z-band Legacy Survey (MzLS; NSF's OIR Lab Proposal ID No. 2016A-0453; PI: Arjun Dey). DECaLS, BASS and MzLS together include data obtained, respectively, at the Blanco telescope, Cerro Tololo Inter-American Observatory, The NSF's National Optical-Infrared Astronomy Research Laboratory (NSF's OIR Lab); the Bok telescope, Steward Observatory, University of Arizona; and the Mayall telescope, Kitt Peak National Observatory, NSF's OIR Lab. The Legacy Surveys project is honored to be permitted to conduct astronomical research on Iolkam Du'ag (Kitt Peak), a mountain with particular significance to the Tohono O'odham Nation.

\noindent The NSF's OIR Lab is operated by the Association of Universities for Research in Astronomy (AURA) under a cooperative agreement with the National Science Foundation.

\noindent This project used data obtained with the Dark Energy Camera (DECam), which was constructed by the Dark Energy Survey (DES) collaboration. Funding for the DES Projects has been provided by the U.S. Department of Energy, the U.S. National Science Foundation, the Ministry of Science and Education of Spain, the Science and Technology Facilities Council of the United Kingdom, the Higher Education Funding Council for England, the National Center for Supercomputing Applications at the University of Illinois at Urbana-Champaign, the Kavli Institute of Cosmological Physics at the University of Chicago, Center for Cosmology and Astro-Particle Physics at the Ohio State University, the Mitchell Institute for Fundamental Physics and Astronomy at Texas A$\&$M University, Financiadora de Estudos e Projetos, Fundacao Carlos Chagas Filho de Amparo, Financiadora de Estudos e Projetos, Fundacao Carlos Chagas Filho de Amparo a Pesquisa do Estado do Rio de Janeiro, Conselho Nacional de Desenvolvimento Cientifico e Tecnologico and the Ministerio da Ciencia, Tecnologia e Inovacao, the Deutsche Forschungsgemeinschaft and the Collaborating Institutions in the Dark Energy Survey. The Collaborating Institutions are Argonne National Laboratory, the University of California at Santa Cruz, the University of Cambridge, Centro de Investigaciones Energeticas, Medioambientales y Tecnologicas-Madrid, the University of Chicago, University College London, the DES-Brazil Consortium, the University of Edinburgh, the Eidgenossische Technische Hochschule (ETH) Zurich, Fermi National Accelerator Laboratory, the University of Illinois at Urbana-Champaign, the Institut de Ciencies de l'Espai (IEEC/CSIC), the Institut de Fisica d'Altes Energies, Lawrence Berkeley National Laboratory, the Ludwig-Maximilians Universitat Munchen and the associated Excellence Cluster Universe, the University of Michigan, the National Optical Astronomy Observatory, the University of Nottingham, the Ohio State University, the University of Pennsylvania, the University of Portsmouth, SLAC National Accelerator Laboratory, Stanford University, the University of Sussex, and Texas A$\&$M University.

\noindent BASS is a key project of the Telescope Access Program (TAP), which has been funded by the National Astronomical Observatories of China, the Chinese Academy of Sciences (the Strategic Priority Research Program "The Emergence of Cosmological Structures" Grant No. XDB09000000), and the Special Fund for Astronomy from the Ministry of Finance. The BASS is also supported by the External Cooperation Program of Chinese Academy of Sciences (Grant No. 114A11KYSB20160057), and Chinese National Natural Science Foundation (Grant No. 11433005).

\noindent The Legacy Survey team makes use of data products from the Near-Earth Object Wide-field Infrared Survey Explorer (NEOWISE), which is a project of the Jet Propulsion Laboratory/California Institute of Technology. NEOWISE is funded by the National Aeronautics and Space Administration.

\noindent The Legacy Surveys imaging of the DESI footprint is supported by the Director, Office of Science, Office of High Energy Physics of the U.S. Department of Energy under Contract No. DE-AC02-05CH1123, by the National Energy Research Scientific Computing Center, a DOE Office of Science User Facility under the same contract; and by the U.S. National Science Foundation, Division of Astronomical Sciences under Contract No. AST-0950945 to NOAO.

\vspace{5mm}
\facilities{{\it Swift}, Magellan:Clay (LDSS), Magellan:Baade (IMACS), Gemini:South (GMOS)}

\software{
IRAF \citep{iraf1, iraf2}, Astrometry.net \citep{Lang_2010}, \texttt{HOTPANTS} \citep{bec15}, \texttt{AstroImageJ} \citep{AstroImageJ}, \texttt{Prospector} \citep{Leja_2017}, \texttt{Python-fsps} \citep{FSPS_2009, FSPS_2010}, \texttt{Dynesty} \citep{Dynesty}, \texttt{lifelines} \citep{lifelines}}

\bibliographystyle{apj}
\bibliography{journals_apj,refs}

\end{document}